\documentclass[12pt]{article}
\usepackage{amsmath}
\usepackage{graphicx,psfrag,epsf}
\usepackage{enumerate}
\usepackage{psfrag,epsf}
\usepackage{url} 
\usepackage{amsmath,amssymb,amsfonts,amsthm,mathtools} 
\usepackage{bm,bbm}
\usepackage{color}
\usepackage{subfigure} 
\usepackage{algorithm,algpseudocode}
\usepackage{scalefnt}
\usepackage{multirow}

\usepackage{natbib}

\newcommand{\blind}{0}

\newtheorem{Remark}{Remark}[section]

\begin{document}

\def\spacingset#1{\renewcommand{\baselinestretch}%
{#1}\small\normalsize} \spacingset{1}


\if0\blind
{
  \title{\bf A parametric quantile beta regression for modeling case fatality rates of COVID-19 }
  \author{  Marcelo Bourguignon\thanks{m.p.bourguignon@gmail.com}\hspace{.2cm}\\
    \small Department of Statistics,  Universidade Federal do Rio Grande do Norte, Natal, Brazil\\
    \\ 
    Diego I. Gallardo\thanks{diego.gallardo.mateluna@gmail.com}\\
   \small Department of Mathematics, Facultad de Ingenier\'ia, Universidad de Atacama, Copiap\'o, Chile\\
    \\
    Helton Saulo\thanks{heltonsaulo@gmail.com}\\
    \small Department of Statistics, University of Bras\'ilia, Bras\'ilia, Brazil
    }
  \maketitle
} \fi

\if1\blind
{
  \bigskip
  \bigskip
  \bigskip
  \begin{center}
    {\Large\bf A parametric quantile beta regression for modeling case fatality rates of COVID-19}
\end{center}
  \medskip
} \fi

\bigskip
\vspace{-0.5cm}
\begin{abstract}
Motivated by the case fatality rate (CFR) of COVID-19, in this paper, we develop a fully parametric quantile regression model based on the generalized three-parameter beta (GB3) distribution. Beta regression models are primarily used to model rates and proportions. However, these models are usually specified in terms of a conditional mean. Therefore, they may be inadequate if the observed response variable follows an asymmetrical distribution, such as CFR data. In addition, beta regression models do not consider the effect of the covariates across the spectrum of the dependent variable, which is possible through the conditional quantile approach. In order to introduce the proposed GB3 regression model, we first reparameterize the GB3 distribution by inserting a quantile parameter and then we develop the new proposed quantile model. We also propose a simple interpretation of the predictor-response relationship in terms of percentage increases/decreases of the quantile.
A Monte Carlo study is carried out for evaluating the performance of the maximum likelihood estimates and
the choice of the link functions. Finally, a real COVID-19 dataset from Chile is analyzed and discussed to illustrate the proposed approach.
\end{abstract}

\noindent%
\vfill

\newpage
\spacingset{1.45} 

\section{Introduction}
\noindent

The new coronavirus respiratory syndrome disease (COVID-19) pandemic has affected several
countries around the world; see \cite{sackeybarfi:2021}. In particular, the COVID-19 pandemic has hit Chile hard in the past few months. Some authors have studied different aspects of COVID-19 in Chile. For instance, \cite{Tariq:21} estimated the reproduction number throughout the epidemic and studied the effectiveness of lockdowns. \cite{Guerrero-Nancuante:20} studied the course of COVID-19 in Chile based on the generalized Susceptible-Exposed-Infectious-Removed model, and \cite{BarriaSandoval:21} discussed different time series methodologies to predict the number of confirmed cases and deaths. In practice, these works can be useful in an epidemic, especially because they provide the basis for making decisions by policy-makers, e.g. directing health care resources to certain areas or identifying how long social distancing policies may need to be in effect.

This work focuses on a measure of COVID-19 mortality risk known as case fatality rate (CFR). Such a measure is particularly important for health policies, and it is computed as the ratio between confirmed deaths and confirmed cases. It is one of
the indicators that serve to monitor the severity of the pandemic; see \url{https://ourworldindata.org/}. In terms of stochastic representation, the CFR can be written as
\begin{equation}\label{ratio1}
\frac{V}{W + V}
\end{equation}
where $V \in \mathbb{R}^+$ and $W \in \mathbb{R}^+$ are two random variables representing the number of confirmed COVID-19-related deaths and COVID-19 cases with no death result, respectively. The sum $W + V$ represents the number
of confirmed COVID-19 cases, and the ratio \eqref{ratio1} has support in the unit interval $(0,1)$. \cite{malik:67}
and \cite{ahuja:69}  both showed that if $V$ and $W$ are independent random variables following
standard (scale parameter equal to 1) gamma distributions with shape parameters $\alpha > 0$ and
$\beta > 0$, i.e. $V \sim \text{GA}(\alpha, 1)$ and $W \sim \text{GA}(\beta, 1)$, then
\begin{equation*}\label{re2}
Y \stackrel{d}{=} \frac{V}{W + V} \sim \text{Beta}(\alpha,\beta),
\end{equation*}
namely, $Y$ is beta distributed with shape parameters $\alpha,\beta > 0$ and $\stackrel{d}{=}$ stands
for equality in distribution.
It should be noted that stochastic representations are important since they may justify some models
arising naturally in real situations, as seen above.

The modeling of CFR of COVID-19 can be approached by regression models that assume a unit interval response, such as the widely used beta regression model. This model was
proposed by \cite{fcn:04}, where the authors consider an alternative parameterization
of the beta distribution in terms of the mean and a precision parameter. In the beta regression model, the mean
is related to a set of covariates through a linear predictor. Nevertheless, it is well known that the widely popular
mean regression model could be inadequate if the observed response variable follows an asymmetrical
distribution (see \cite{galarzaetal:17} and \cite{slgs:2020a}), which is quite common for rates and proportions.
In such a situation, quantile regression models
\citep{k:05} can be more suitable. Quantile regression is a methodology used for understanding
the conditional distribution of a response variable given the values of some covariates at different
levels (quantiles), thus providing users with a more complete picture.
Despite this, as far as we know, a specific
parametric quantile regression model to describe data observed in the unitary interval at different quantiles that satisfies the stochastic representation in \eqref{ratio1} has never been considered in the literature.
This paper works in this direction.


Although the beta regression is the standard model for quantifying the influence of covariates on the mean of the response variable in the unit interval, it is not useful for quantifying such an influence on the quantiles of the dependent variable. This occurs because the cumulative distribution function (CDF) of the beta distribution does not have an invertible closed form, which hinders its utilization for parametric quantile regression purposes.

A natural approach to model CFR in observational studies is to apply a logarithmic transformation to the
variable $Y/(1-Y)$ and fit a regression model based on the log-transformed variable.
In this context, \cite{bottai10} introduced a logistic quantile regression for the analysis of bounded outcomes.
In particular, the authors studied a standardized logistic
transformation for an arbitrary interval response.
Thus, within this approach, data are analyzed on a non-original scale, which
complicates the interpretation of the parameters in applied studies.
Additionally, the methodology developed by \cite{bottai10} requires the addition of a small quantity to the bounds to ensure that the logistic transformation is defined for all values of the response. Moreover, it is only possible to obtain a regression structure for the quantiles,
so the model is not flexible. Finally, this approach lacks interpretation in terms of the CFR quotient presented in \eqref{ratio1}.

Recently, \cite{Smithson17} developed a CDF-quantile family of two-parameter
distributions with support in $(0, 1)$. Such family enables a wide variety of quantile regression models for random variables on
the unit interval, with predictors for both the location and scale parameters. However, given a dataset, the authors do not make it clear how many CDF-quantile models should be tried in
practice, and what should be done when there is no significant difference among different fitted CDF-quantile models.
An alternative to the beta distribution is the Kumaraswamy distribution.
\cite{bayesetal:17} studied a quantile parametric mixed regression model for bounded
response variables considering the Kumaraswamy distribution. However, this distribution, as far as we know, does not present the  stochastic representation in (\ref{ratio1}).
Such criticism can be extended to the CDF-quantile family of two-parameter distributions \citep{Smithson17}.

On the other hand, a probability distribution related to the beta model that has received little attention in
the literature was introduced by \cite{libbynovick:82}, named the Libby-Novick beta distribution
or generalized beta with three parameters (GB3) distribution. This distribution is a generalization of the
standard beta distribution and can be an interesting and useful model for modeling double bounded data satisfying \eqref{ratio1}, which is the main feature of CFR data. The GB3 distribution is a three-parameter ge\-ne\-ra\-li\-za\-tion
of the usual two-parameter beta distribution, and it is perhaps the simplest generalization of the
two-parameter beta distribution that allows for significant additional flexibility; see
\cite{risticetal:15}. The GB3 distribution offers more flexibility for modeling real data
than the beta distribution because the additional shape parameter can control the skewness and
kurtosis simultaneously, varying tail weights and providing more entropy; see
\cite{cordeirosantanaetal:14}. Furthermore, as said earlier, the GB3 distribution presents the stochastic representation in \eqref{ratio1}. Despite all these characteristics, little attention
in the statistical literature has been devoted to the GB3 distribution.

In this work, motivated by the features of COVID-19 CFR data, we propose and study a new parametric quantile
regression model based on the GB3 distribution and verify that the proposed quantile regression model is suitable for modeling double bounded data satisfying the stochastic representation in (\ref{ratio1}). To the
best of our knowledge, a specific parametric
quantile regression model to describe data of the type $V/(W + V)$ at different levels
has never been considered in the literature. The quantile approach allows for capturing the influence of the covariates on the spectrum of the dependent variable, in addition to coping with outliers; see
\cite{k:05}, \cite{hn:07}, and \cite{waldmann:18}.
Furthermore, the proposed quantile regression model allows for the quantiles of the data to be described on their
original scale, unlike the existing models that employ a logarithmic transformation of the CFR.
Similarly to the works of \cite{nj:13} and \cite{ssls:21}, who introduced quantile
parameters in the generalized gamma and log-symmetric distributions, respectively, we introduce a reparameterization of the GB3 model by inserting a quantile
parameter before developing the GB3 quantile
regression model. Our proposed approach relates the quantiles of \eqref{ratio1}, $q(V/(W + V)|\mathbf{X})$,
to a set of explanatory variables $\mathbf{X} = (X_{1}, \ldots, X_{k})$, providing a full parametric elegant quantile
regression model; see \cite{bergeretal:19} for a similar approach related to the ratio of two positively correlated
biomarkers. We also propose a simple interpretation of the predictor-response relationship in terms of percentage
increases/decreases of the quantile. We analyze COVID-19 data from Chile and find that the CFR is explained by population density, positivity for tests, and the percentage of the population fully vaccinated.

The rest of the paper unfolds as follows. In Section \ref{sec:2}, we describe the classical GB3 distribution and propose a reparameterization of this distribution in terms of a quantile parameter. In Section \ref{sec:3}, we introduce the GB3 quantile regression model and discuss the estimation of the model parameters by the maximum likelihood (ML) method. We also discuss residual analysis and covariate selection in this section. In Section~\ref{sec:4}, we carry out a Monte Carlo simulation study to evaluate the recovery of the parameters and assess the choice of the link functions. In Section \ref{sec:5}, we apply the GB3 quantile regression model to a real COVID-19 CFR dataset. Finally, in Section \ref{sec:6}, we provide some concluding remarks.

\section{Preliminaries}\label{sec:2}
\noindent

In this section, we shall describe the GB3 distribution and introduce the quantile-based reparameterization of this distribution, which will be useful subsequently for developing the GB3 quantile regression model.

\subsection{The classical GB3 distribution}\label{sec:2.1}
\noindent

A random variable $Y$ follows the generalized beta distribution
with three parameters, $\alpha > 0, \beta > 0$, and $\lambda > 0$, denoted by
$Y \sim \textrm{GB3}(\lambda, \alpha, \beta)$, if its CDF is given by
\begin{equation}\label{int:01}
F_Y(y; \lambda, \alpha, \beta) = I_{\lambda\,x/(1+\lambda\,x-x)}(\alpha, \beta),
 \quad 0 < y < 1,
\end{equation}
where $I_{x}(\alpha, \beta) = B_x(\alpha, \beta)/B(\alpha, \beta)$ is the incomplete beta function ratio,
$B_x(\alpha, \beta) = \int_{0}^{x}\omega^{\alpha-1}(1-\omega)^{\beta-1}\textrm{d} \omega$ is the incomplete function and $B(\alpha, \beta)=B_1(\alpha,\beta)$ is the usual beta function.

The probability density function (PDF) associated with Equation \eqref{int:01} is
\begin{equation}\label{inv:01}
f_Y(y; \lambda, \alpha, \beta) = \frac{\lambda^\alpha y^{\alpha-1}(1 - y)^{\beta-1}}{B(\alpha, \beta)[1-(1-\lambda)y]^{\alpha+\beta}}, \quad 0 < y < 1.
\end{equation}
We can see that Equation (\ref{inv:01}) reduces to the beta distribution when $\lambda = 1$. If $Y \sim \textrm{GB3}(\lambda, \alpha, \beta)$,
then its complement $(1 - Y)$ is $\textrm{GB3}(\lambda^{-1}, \beta, \alpha)$ distributed. This property is similar to the one enjoyed by
the beta distribution. According to \cite{cordeirosantanaetal:14}, taking $\alpha = \beta$, the shape parameter $\lambda$ gives tail weights of the PDF to the
extreme right, and skewness and kurtosis increase when $\lambda$ approaches zero for
$0 < \lambda < 1$. On the other hand, the PDF \eqref{inv:01} becomes more symmetric when $\lambda$ approaches one,
and for $\lambda > 1$, the parameter $\lambda$ gives tail weights of the PDF to the
extreme left when $\lambda \rightarrow \infty$. Furthermore, the skewness and kurtosis increase when $\lambda$ tends to infinity.

\begin{Remark}
(Relation to the generalized beta distribution of the first kind model)
The generalized beta distribution of the first kind \citep{MC95}, GB1($\mu, \sigma, \nu, \tau$), is defined by the following PDF:
\begin{equation}\label{gb1}
f_Y(y; \mu, \sigma, \nu, \tau) = \frac{\tau\,\nu^{\beta}y^{\tau\,\alpha-1}(1-y^\tau)^{\beta-1}}
{B(\alpha, \beta)[\nu + (1-\nu)y^\tau]^{\alpha + \beta}}, \quad 0 < y < 1,
\end{equation}
where $\alpha = \mu(1 - \sigma^2)/\sigma^2 > 0$ and $\beta = (1-\mu)(1-\sigma^2)/\sigma^2 > 0$; see \cite{sta07}.
Comparing with (\ref{inv:01}), we can see that the GB3$(\alpha, \beta, \lambda)$ model
is a special type of the GB1($\mu, \sigma, \nu, \tau$) model as follows:
$$\textrm{GB3}(\lambda, \alpha, \beta) =
\textrm{GB1}(\alpha(\alpha + \beta)^{-1}, (\alpha + \beta - 1)^{-1/2}, \lambda^{-1}, 1).$$

The GB1 model was introduced by \cite{MC95} as an income
distribution and additional details can
be found in their paper.
\end{Remark}

In a similar way to the classical beta distribution, if $X_1 \sim \mathrm{GA}(\alpha, \theta_1)$ and $X_2 \sim \mathrm{GA}(\beta, \theta_2)$ are independent gamma distributions, then the random variable \citep{libbynovick:82}
\begin{equation}\label{re1}
Y \stackrel{d}{=} \frac{X_1}{X_1 + X_2} \sim \mathrm{GB3}(\alpha, \beta, \lambda)
\end{equation}
is distributed according to a GB3 distribution, where $\lambda = \theta_1/\theta_2$.



Note that, as $I_{x}(\alpha,\beta)$ corresponds to the CDF of the usual beta distribution,
the $\tau$-th quantile of the GB3 distribution can be represented as
\begin{equation}\label{quant}
q(\tau; \lambda, \alpha, \beta) = \frac{z_{\alpha,\beta}(\tau)}{\lambda [1-z_{\alpha,\beta}(\tau)] + z_{\alpha,\beta}(\tau)}, \quad 0 < \tau < 1,
\end{equation}
where $z_{\alpha,\beta}(\tau)$ denotes the $\tau$-th quantile of the beta distribution with parameters $\alpha$ and $\beta$.
To compute the quantile $z_{\alpha,\beta}(\tau)$ as required, it is not necessary to
implement such complex formulae as (\ref{int:01}). Instead, one can use the $\texttt{qbeta}(\cdot)$ function in \texttt{R} \citep{R:21}.

\subsection{The quantile-based GB3 distribution}\label{sec:2.2}
\noindent

To introduce the proposed quantile regression model, we shall reparametrize \eqref{inv:01} in
terms of the $\tau$-th quantile $\mu = q(\tau; \lambda, \alpha, \beta)$ such that $\lambda$ can be written (from Eq. (\ref{quant})) as follows
\begin{equation*}
\lambda = \frac{(1-\mu)}{\mu} \frac{z_{\alpha,\beta}(\tau)}{[1-z_{\alpha,\beta}(\tau)]}, \quad 0 < \mu < 1.
\end{equation*}

\begin{figure}[!htbp]
\vspace{-0.30cm}
\centering
\vspace{-0.15cm}\subfigure[$\tau=0.1$]{\includegraphics[height=5.0cm,width=4.0cm]{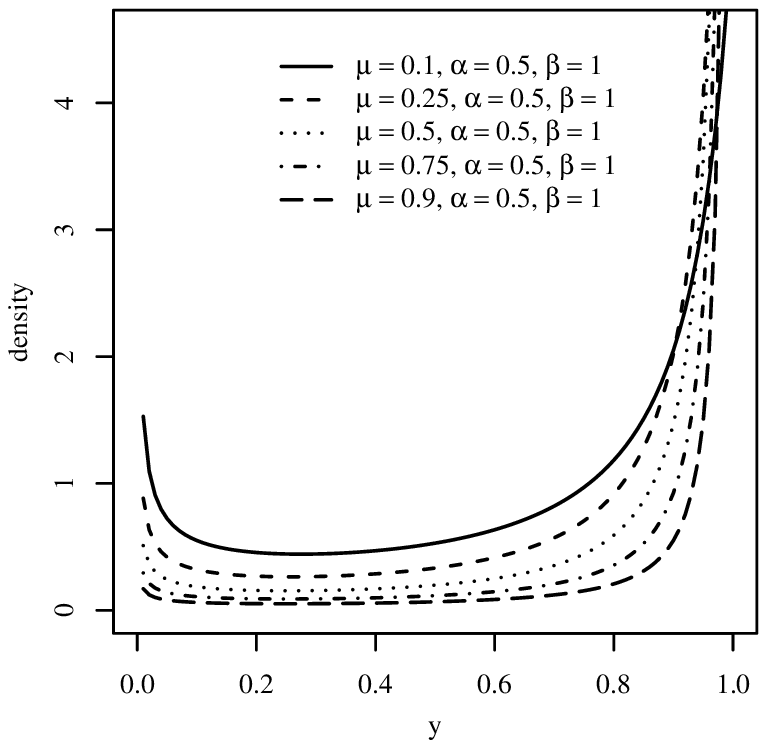}}
\vspace{-0.15cm}\subfigure[$\tau=0.5$]{\includegraphics[height=5.0cm,width=4.0cm]{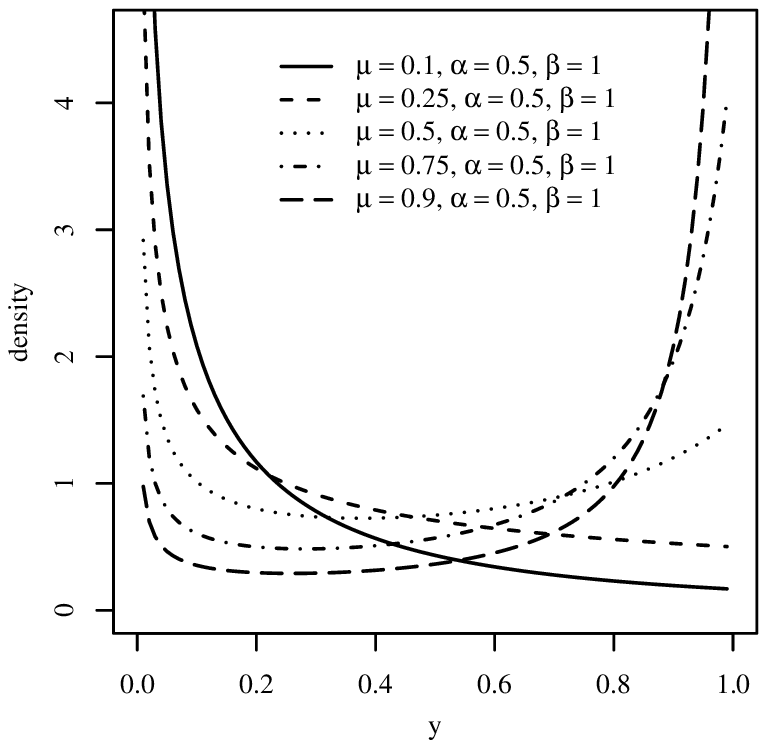}}
\vspace{-0.15cm}\subfigure[$\tau=0.9$]{\includegraphics[height=5.0cm,width=4.0cm]{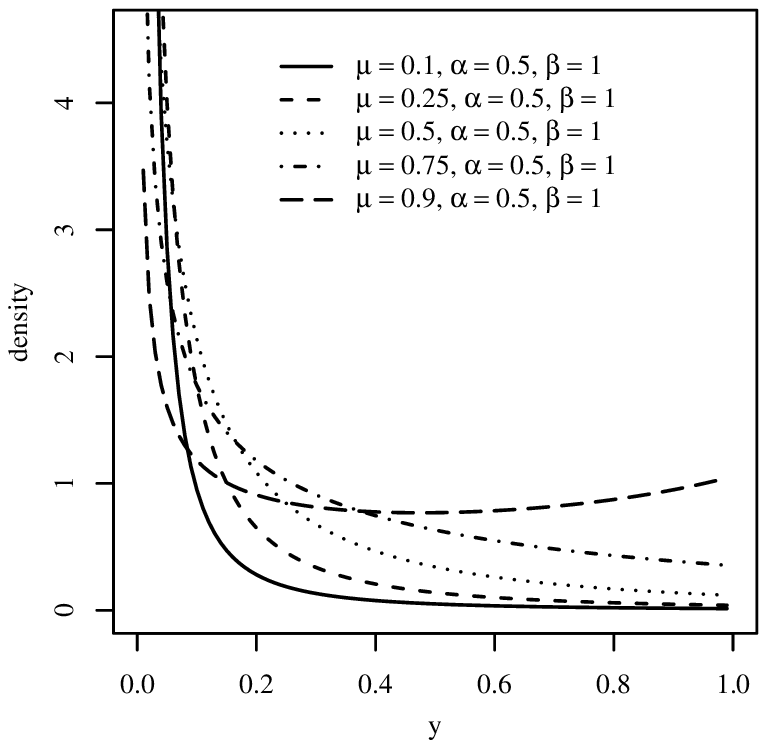}}\\
\vspace{-0.15cm}\subfigure[$\tau=0.1$]{\includegraphics[height=5.0cm,width=4.0cm]{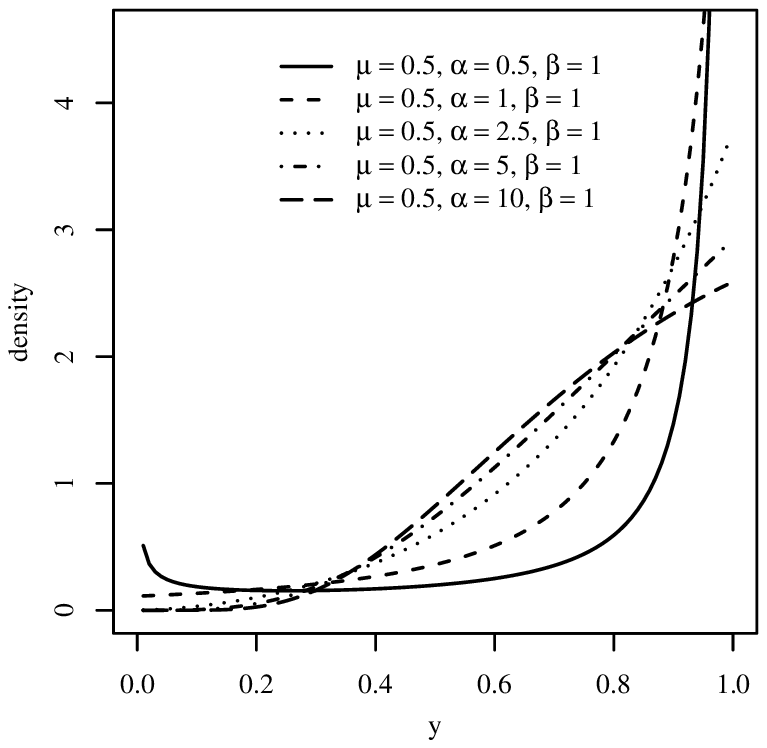}}
\vspace{-0.15cm}\subfigure[$\tau=0.5$]{\includegraphics[height=5.0cm,width=4.0cm]{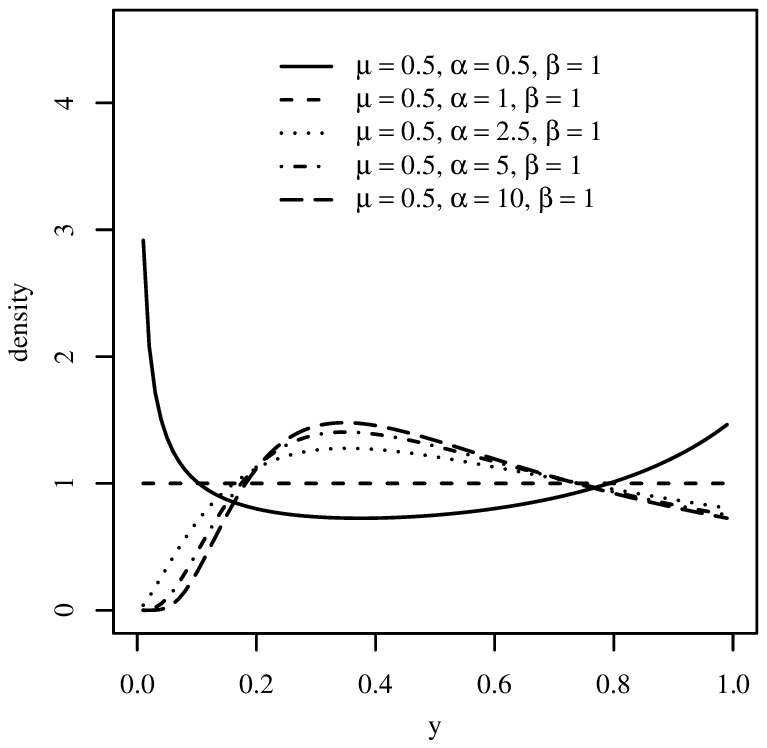}}
\vspace{-0.15cm}\subfigure[$\tau=0.9$]{\includegraphics[height=5.0cm,width=4.0cm]{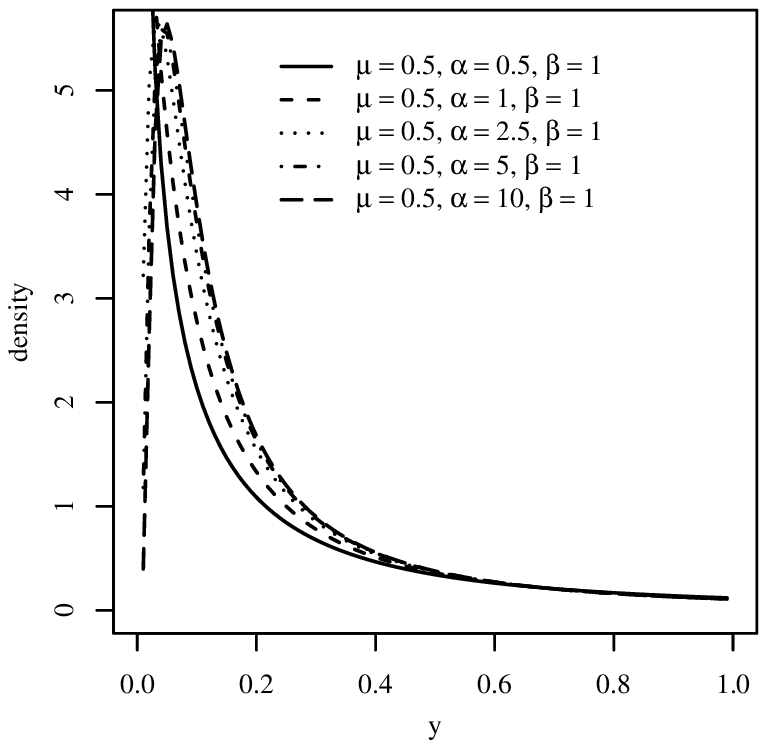}}\\
\vspace{-0.15cm}\subfigure[$\tau=0.1$]{\includegraphics[height=5.0cm,width=4.0cm]{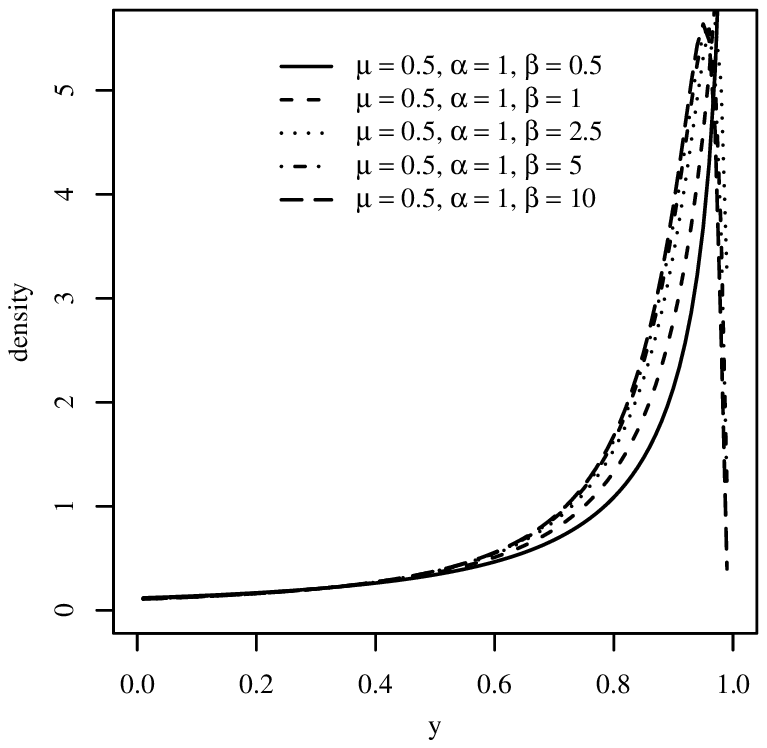}}
\vspace{-0.15cm}\subfigure[$\tau=0.5$]{\includegraphics[height=5.0cm,width=4.0cm]{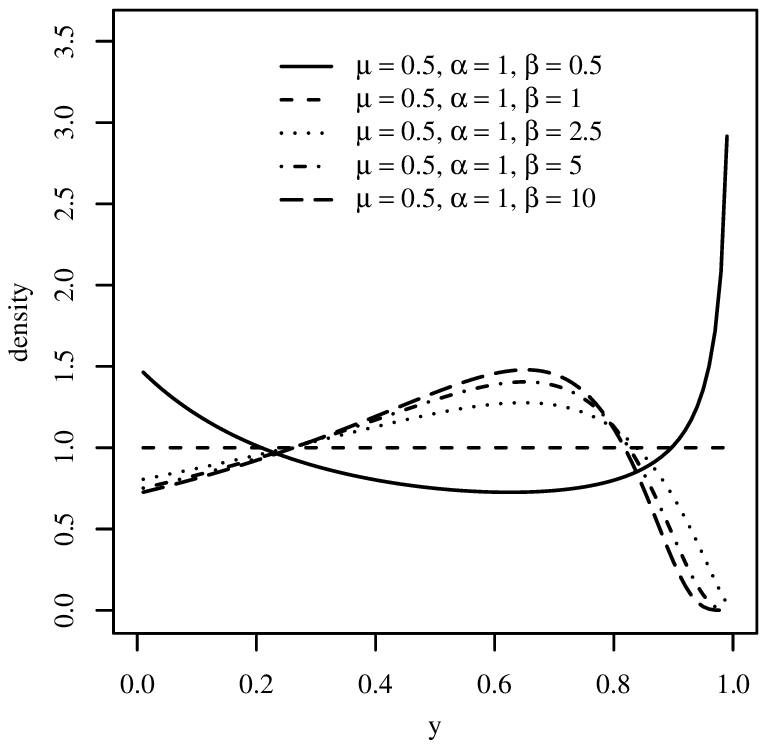}}
\vspace{-0.15cm}\subfigure[$\tau=0.9$]{\includegraphics[height=5.0cm,width=4.0cm]{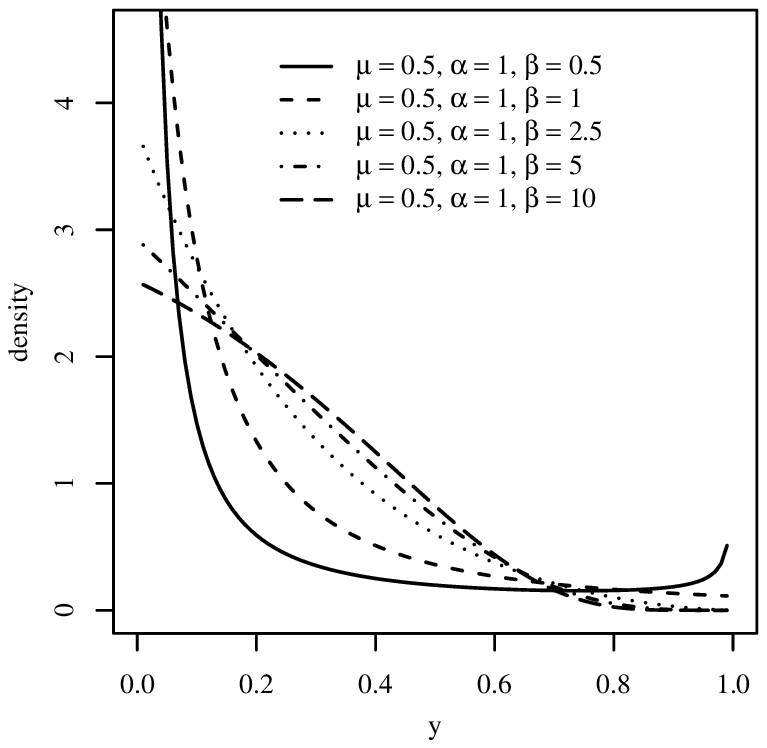}}
\caption{GB3 PDFs for some combinations of parameter values.}\label{figpdfs}
\end{figure}

This parametrization has not been proposed in the statistical literature. Under this new parametrization,
the PDF and CDF of the GB3 distribution can be written, respectively, as
\begin{equation*}\label{int:02}
F_Y(y; \mu, \alpha, \beta) = I_{\frac{x(1-\mu)}{\mu} \frac{z_{\alpha,\beta}(\tau)}{[1-z_{\alpha,\beta}(\tau)]}/\left(1+\frac{(1-\mu)}{\mu} \frac{z_{\alpha,\beta}(\tau)}{[1-z_{\alpha,\beta}(\tau)]}x-x\right)}\left(\alpha, \beta\right)
\end{equation*}
and
\begin{equation}\label{inv:02}
f_Y(y; \mu, \alpha, \beta) = \frac{\left(\frac{(1-\mu)}{\mu} \frac{z_{\alpha,\beta}(\tau)}{[1-z_{\alpha,\beta}(\tau)]}\right)^\alpha y^{\alpha-1}(1 - y)^{\beta-1}}{B(\alpha, \beta)\left[1-\left(1-\frac{(1-\mu)}{\mu} \frac{z_{\alpha,\beta}(\tau)}{[1-z_{\alpha,\beta}(\tau)]}\right)y\right]^{\alpha+\beta}}, \quad 0 < y < 1.
\end{equation}
Hereafter, we shall use the notation $Y \sim \textrm{GB3}(\mu, \alpha, \beta)$, where $\mu \in (0, 1)$ is the quantile
parameter, $\alpha > 0$ and $\beta > 0$ are the shape parameters, and $\tau \in (0, 1)$ is assumed to be known. Figure \ref{figpdfs} displays different shapes of the GB3 PDF for different combinations of parameters.
It is noteworthy that the GB3 is very flexible since its density assumes different forms.

\section{The GB3 quantile regression model}\label{sec:3}
\noindent

In this section, we shall introduce the GB3 quantile regression model and discuss parameter estimation based on the ML method. We also consider residuals analysis and covariate selection for the proposed model.

\subsection{The model}\label{sec:3-1}
\noindent

Let $Y_1, \ldots, Y_n $ be independent random
variables such that each $ Y_i $, for $ i = 1,\ldots,n$, has PDF defined in \eqref{inv:02}, $Y_i \sim \textrm{GB3}(\mu_i(\tau), \alpha_i(\tau), \beta_i(\tau))$, for a fixed (known) probability $
\tau \in (0, 1)$ associated with the quantile of interest.
Suppose $\mu(\tau)$, $\alpha(\tau)$, and $\beta(\tau)$ satisfy the following functional relations:
\begin{equation}\label{cs1}
g_1(\mu_i(\tau)) = \mathbf{x}^\top_{i}\bm{\theta}(\tau), \quad g_2(\alpha_i(\tau)) = \mathbf{z}^\top_{i}\bm{\nu}(\tau), \quad \textrm{and} \quad g_3(\beta_i(\tau))  = \mathbf{w}^\top_{i}\bm{\eta}(\tau),
\end{equation}
where $\bm{\theta}(\tau) = (\theta_{0}(\tau), \ldots, \theta_{k}(\tau))^\top$,  $\bm{\nu}(\tau) = (\nu_{1}(\tau), \ldots, \nu_{l}(\tau))^\top$, and  $\bm{\eta}(\tau) = (\eta_{1}(\tau), \ldots, \eta_{m}(\tau))^\top$
are the vectors of the unknown regression coefficients, which are assumed to be functionally independent;
$\bm{\theta}(\tau) \in \mathbb{R}^k$, $\bm{\nu}(\tau) \in \mathbb{R}^l$, and $\bm{\eta}(\tau) \in \mathbb{R}^m$, with $k + l + m < n$;
and $\mathbf{x}_{i} = (x_{i1}, \ldots, x_{ik})^\top$, $\mathbf{z}_{i} = (z_{i1}, \ldots, z_{il})^\top$, and $\mathbf{w}_{i} = (w_{i1}, \ldots, w_{im})^\top$ are the observations of the $k, l$, and $m$ known regressors, for $i = 1, \ldots, n$.
Furthermore, we assume that the covariate matrices $\mathbf{X} = (\mathbf{x}_1, \ldots, \mathbf{x}_n)^\top$, $\mathbf{Z} = (\mathbf{z}_1, \ldots, \mathbf{z}_n)^\top$, and $\mathbf{W} = (\mathbf{w}_1, \ldots, \mathbf{w}_n)^\top$ have ranks $k, l$, and $m$, respectively. The link functions $g_1: (0, 1) \rightarrow \mathbb{R}$, $g_2: \mathbb{R}^+ \rightarrow \mathbb{R}$, and $g_3: \mathbb{R}^+ \rightarrow \mathbb{R}$ in (\ref{cs1}) must be strictly monotone, positive, and at least twice differentiable, with $g_1^{-1}(\cdot)$, $g_2^{-1}(\cdot)$, and $g_3^{-1}(\cdot)$ being the inverse functions of $g_1(\cdot)$, $g_2(\cdot)$, and $g_3(\cdot)$, respectively. For $g_1(\cdot)$, the most common choices are the logit, probit, loglog, and complementary loglog (cloglog) link functions, whereas for $g_2(\cdot)$
and $g_3(\cdot)$ the most common choice is the log link.

\subsection{Interpretation of the GB3 quantile model}
\noindent

The interpretation of the regression coefficients related to the quantile $\mu_i$ can be done in terms of percentage change. Consider ${\theta}_j(\tau)$ as the $j$-th regression coefficient where $j$ denotes the exclusion of the $j$-th element. Then, $\mathbf{x}_{i(j)}$ and ${\bm{\theta}_{(j)}(\tau)}$ denote the vector of covariates excluding $x_{ij}$ and the vector of regression coefficients excluding ${\theta}_j(\tau)$, respectively. Note that when $x_{ij}$ increases by one, holding $\mathbf{x}_{i(j)}$ fixed, we obtain
\begin{eqnarray}\label{intercoeffirst}\nonumber
 \mu_i({x}_{ij}+1,\mathbf{x}_{i(j)})&=& g_{1}^{-1}({\theta}_j(\tau)({x}_{ij}+1)+ \mathbf{x}^\top_{i(j)}{\bm{\theta}_{(j)}(\tau)})\\ \nonumber
                                  &=&g_{1}^{-1}({\theta}_j(\tau)+{\theta}_j(\tau){x}_{ij}+ \mathbf{x}^\top_{i(j)}{\bm{\theta}_{(j)}(\tau)})\\
                                  &=&g_{1}^{-1}({\theta}_j(\tau) + \mathbf{x}^\top_{i}\bm{\theta}(\tau)  ).
\end{eqnarray}
Therefore, for any $j$ increasing $x_{ij}$ by one unit, the quantile $\mu_i$ can be expressed as a percentage change as follows:
\begin{equation}\label{intercoeff}
 \frac{ \mu_i({x}_{ij}+1,\mathbf{x}_{i(j)}) - \mu_i({x}_{ij},\mathbf{x}_{i(j)}) }{ \mu_i({x}_{ij},\mathbf{x}_{i(j)}) }\times 100\%
 = \frac{g_{1}^{-1}({\theta}_j(\tau) + \mathbf{x}^\top_{i}\bm{\theta}(\tau))-g_{1}^{-1}(\mathbf{x}^\top_{i}\bm{\theta}(\tau))}{g_{1}^{-1}(\mathbf{x}^\top_{i}\bm{\theta}(\tau))}
 \times 100\%.
\end{equation}
That is, \eqref{intercoeff} provides the percentage increase (or decrease if the value of $\beta_j$ is negative) in the quantile $\mu_i$ when ${x}_{ij}$ is increased by one unit. For $x_{ij}$ dichotomous, \eqref{intercoeff} provides the percentage increase (or decrease if $\beta_j$ is negative) in the quantile $\mu_i$ when ${x}_{ij}$ changes from category 0 to 1; see \cite{weisberg:14}. Note that when ${x}_{i}=\log({z}_{i})$, ${z}_{i}$ can be replaced by $c\,{z}_{i}$, and \eqref{intercoeffirst} can be rewritten as
\begin{eqnarray}\label{intercoeffirst2}\nonumber
 \mu_i({x}_{ij}+\log(c),\mathbf{x}_{i(j)})&=& g_{1}^{-1}({\theta}_j(\tau)({x}_{ij}+\log(c))+ \mathbf{x}^\top_{i(j)}{\bm{\theta}_{(j)}(\tau)})\\ \nonumber
                                  &=&g_{1}^{-1}({\theta}_j(\tau)\log(c)+{\theta}_j(\tau){x}_{ij}+ \mathbf{x}^\top_{i(j)}{\bm{\theta}_{(j)}(\tau)})\\
                                  &=&g_{1}^{-1}({\theta}_j(\tau)\log(c) + \mathbf{x}^\top_{i}\bm{\theta}(\tau)  ).
\end{eqnarray}
For example, if ${z}_{i}$ is increased by $5\%$, then $c=1.05$ and the percentage change \eqref{intercoeff} is computed taking into account \eqref{intercoeffirst2} with $\log(1.05)\approx 0.5$. However, this interpretation not only depends on ${\theta}_j(\tau)$
but also on the remaining coefficients.

To provide an alternative and more useful interpretation, we can use a Taylor series expansion of first order to approximate $\log[ g^{-1}(x)]$ to zero. For the cases where $g_1(\cdot)$ corresponds to the logit, probit, loglog, and cloglog links, we have that $\log[g^{-1}(x)]\approx a_0+a_1 x$, where $a_0=-\log (2)$ and $a_1=1/2$ for logit; $a_0=-\log (2)$ and $a_1=\sqrt{2/\pi}$ for probit; $a_0=-\textrm{e}$ and $a_1=1$ for loglog; and $a_0=\log(1-\exp(-\textrm{e}))$ and $a_1=[1+\exp(-\textrm{e})]^{-1}-1$ for cloglog. Therefore,
\begin{eqnarray*}
\log[\mu_i({x}_{ij}+1,\mathbf{x}_{i(j)})]-\log[ \mu_i({x}_{ij}],\mathbf{x}_{i(j)})&\approx& a_0 +a_1 \left[{\theta}_j(\tau)+ \mathbf{x}^\top_{i}{\bm{\theta}(\tau)}\right]-a_0 \\ &-&a_1 \mathbf{x}^\top_{i}{\bm{\theta}(\tau)}=a_1{\theta}_j(\tau)
\end{eqnarray*}
and
\begin{align}
\frac{\mu_i({x}_{ij}+1,\mathbf{x}_{i(j)})-\mu_i({x}_{ij},\mathbf{x}_{i(j)})}{\mu_i({x}_{ij},\mathbf{x}_{i(j)})}\approx \exp\left[-a_1{\theta}_j(\tau)\right]-1, \label{interp}
\end{align}
which only depends on ${\theta}_j(\tau)$. For this reason, $100\times \left(\exp\left[-a_1{\theta}_j(\tau)\right]-1\right)\%$ can be interpreted as the  approximate percentage increase (or decrease) in the response variable when the $j$-th covariate related to the quantile is increased by one unit and the remaining covariates are fixed.

Additionally, by the stochastic representation in Equation (\ref{re1}), we have that, conditional on $\mathbf{z}^\top_{i}$ and $\mathbf{w}^\top_{i}$, the latent variables $X_1$ and $X_2$ are GA$(\alpha_i(\tau),\theta_1)$ and GA$(\beta_i(\tau),\theta_2)$ distributed, respectively. It is well known that the means of such distributions are $\alpha_i(\tau)/\theta_1$ and $\beta_i(\tau)/\theta_2$, respectively. Therefore, an increment of one unit in the $j$-th covariate of $\mathbf{z}^\top_i$ produces the following relative change in the mean of $X_1$:
\begin{align*}
\frac{E(X_1\mid z_{ij}+1,\mathbf{z}_{i(j)})-E(X_1\mid z_{ij},\mathbf{z}_{i(j)})}{E(X_1\mid z_{ij},\mathbf{z}_{i(j)})}&=\frac{\exp(\nu_{j}(\tau)+\mathbf{z}_i^\top {\bm \nu }(\tau))/\theta_1-\exp(\mathbf{z}_i^\top {\bm \nu}(\tau))/\theta_1}{\exp(\mathbf{z}_i^\top {\bm \nu}(\tau))/\theta_1}\\
&~~=\exp(\nu_j(\tau))-1.
\end{align*}
In other words, $100 \times \left[\exp(\nu_j(\tau))-1\right] \%$ represents the relative increment (or decrement) in the mean of $X_1$ when the $j$-th covariate increases one unit and the rest of the covariates are maintained as fixed. Similarly, it is possible to show that $100 \times \left[\exp(\eta_j(\tau))-1\right] \%$ represents the increment/decrement in the mean of $X_2$ when the $j$-th covariate of $\mathbf{w}_i^\top$ increases one unit and the rest of the covariates are maintained. In the CFR of COVID-19, this allows for the interpretation of the relative changes directly in terms of the number of deaths and cases of COVID-19 for the population density, average of positivity for the tests in a determined past period, and percentage of the population fully vaccinated in Chile, for instance. This interpretation is valid for any $\tau \in (0,1)$. Note that it is possible that a covariate could be related to the total deaths and/or total cases of COVID-19, but not related to the quantile of the CFR of COVID-19, and vice versa.
Henceforth, in order to avoid overloading the notation, we omit $\tau$ in $\bm{\theta}, \bm{\nu}$ and $\bm{\eta}$ and in $\mu_i, \alpha_i$ and $\beta_i$. 

\subsection{Estimation and inference}\label{ML}
\noindent

Let $Y_1, \ldots, Y_n$ be a set of $n$ independent random variables such that $Y_i \sim \textrm{GB3}(\mu_i, \alpha_i, \beta_i)$, and $y_1,\ldots,y_n$ be the corresponding observations. The corresponding likelihood function for
$\bm{\Theta} = (\bm{\theta}^{\top},\bm{\nu}^{\top},\bm{\eta}^{\top})^{\top}$ is
\begin{equation}\label{eq:loglik1}
L({\bm\Theta})= \prod_{i=1}^{n} \frac{\left(\frac{(1-\mu_i)}{\mu_i} \zeta_{i}\right)^{\alpha_i} y_i^{\alpha_i-1}(1 - y_i)^{\beta_i-1}}{B(\alpha_i, \beta_i)\left[1-\left(1-\frac{(1-\mu_i)}{\mu_i} \zeta_i\right)y_i\right]^{\alpha_i+\beta_i}} ,
\end{equation}
where $\mu_i$, $\alpha_i$, and $\beta_i$ are given in \eqref{cs1} and
\begin{equation*}
\zeta_{i}=\frac{z_{\alpha_i,\beta_i}(\tau)}{[1-z_{\alpha_i,\beta_i}(\tau)]}.
\end{equation*}

By taking the logarithm of \eqref{eq:loglik1}, we obtain the log-likelihood function for $\bm{\Theta}$,
\begin{equation}\label{eq:loglik2}
\ell({\bm\Theta}) = \sum_{i=1}^{n} \ell_i({\bm\Theta}),
\end{equation}
where
\begin{eqnarray*}
\ell_i({\bm\Theta}) &=& {\alpha_i}\log(1-\mu_i)-\alpha_i\log(\mu_i) +{\alpha_i}\log(\zeta_{i})  + (\alpha_i-1)\log(y_i)\\&+& (\beta_i-1)\log(1 - y_i)-\log\left( B(\alpha_i, \beta_i) \right) \\&-& (\alpha_i+\beta_i) \log\left(1-\left(1-\frac{(1-\mu_i)}{\mu_i} \zeta_i\right)y_i\right).
\end{eqnarray*}
The score vector $\dot{\bm \ell}({\bm\Theta})$ is obtained by differentiating the log-likelihood function in \eqref{eq:loglik2} with respect to each component of $\bm{\Theta}$. By equating the score vector to zero, we obtain the ML estimate of ${\bm\Theta}$. However, as the ML estimators of $\bm{\Theta}$ do not have closed form
expressions, the parameters must be estimated using standard numerical optimization algorithms.
We use the Broyden-Fletcher-Goldfarb-Shanno (BFGS) quasi-Newton method; see \citet[][p.\,199]{mjm:00}.
Numerical maximization of the log-likelihood function is accomplished by using the \texttt{R} software (available at \url{http://www.r-project.org}). The computational program is available from authors upon request.

Under mild regularity conditions \citep{Cox1979} and when $ n $ is large,
the asymptotic distribution of the ML estimators
$\widehat{\bm{\Theta}} = (\widehat{\bm{\theta}}^{\top},\widehat{\bm{\nu}^\top},\widehat{\bm{\eta}^\top})^{\top}$
is approximately multivariate normal (of dimension $k + l + m$) with mean vector
$\bm{\Theta} = (\bm{\theta}^{\top},\bm{\nu}^\top,\bm{\eta}^\top)^{\top}$  and variance-covariance matrix
$\mathbf{K}^{-1}(\bm{\Theta})$, where
$$\mathbf{K}(\bm{\Theta})= \mathbb{E}\left[- \ {\partial \ell \left(\bm{\Theta}\right)\over \partial \bm{\Theta} \; \partial \bm{\Theta}^\top} \right]$$
is the  expected Fisher information matrix.
Unfortunately, there is no closed form expression for $\mathbf{K}(\bm{\Theta})$.
Nevertheless, a consistent estimator of the expected Fisher information matrix is given by
$$\mathbf{J}(\widehat{\bm{\Theta}})=- \ {\partial \ell \left(\bm{\Theta}\right)\over \partial \bm{\Theta} \;\partial \bm{\Theta}^\top} \Big{|}_{\bm{\Theta} = \widehat{\bm{\Theta}}} \ ,$$
which is the estimated observed Fisher information matrix.
Therefore, for large $n$, we can approximate $\mathbf{K}(\bm{\Theta})$ by $\mathbf{J}(\widehat{\bm{\Theta}})$.

\subsection{Model adequacy}
\noindent

To assess the goodness of fit and departures from the assumptions of the model, we shall use the randomized quantile \citep{ds:96} (RQ) residual given by
\begin{equation*}\label{eq_sec:randquantile}
 R^\textrm{RQ}_i = \Phi^{-1}\big(\widehat{F}_Y(y_i; \widehat{\mu}_i, \widehat{\alpha}_i, \widehat{\beta}_i)\big),\quad i=1, \ldots, n,
\end{equation*}
where $\Phi^{-1}$ is the inverse function of the standard normal CDF and $\widehat{F}(\cdot)$ is the CDF fitted to the data. This residual follows approximately a standard normal distribution when the model is correctly specified. Then, we can use quantile-quantile (QQ) plots to assess the fit.

\subsection{Covariate selection}\label{cov.selection}
\noindent

A critical point of the model is the selection of the covariates included in each component of the model: $\mu_i, \alpha_i$, and $\beta_i$.
If we dispose of $r$ covariates for each individual and considering that the intercept is included in all components, we can arrange such covariates in $2^{3r}$ ways, i.e., such quantity of models. A way to select a combination of covariates is to fit the $2^{3r}$ models and select the one with minimum Akaike information criteria \citep[AIC; ][]{akaike1974} or Bayesian information criteria \citep[BIC; ][]{s:78}, for instance. However, such method demands a high computational cost and some covariates could not be significant for some significance levels. To avoid these problems, an alternative is to fit the model with all covariates in the three components and successively eliminate the covariate (which can be in any of the 3 components of the model) with the lower significance (i.e., the greater $p$-value based on the asymptotic distribution of the ML estimator). Under this scheme, we compute the estimates for $3r$ models at maximum.

\section{Simulation studies}\label{sec:4}
\noindent

In this section, we present two simulation studies to assess different aspects of the GB3 regression model. The first study assesses the recovery parameters of the model under different combinations of sample size and $\tau$. The second study evaluates the performance of the model selection using information criteria to select a determined link to the quantile parameter in the GB3 regression model. For all cases, we considered the design matrices $\mathbf{x}_i=(1,t_{1i},t_{2i})$, $\mathbf{z}_i=(1,t_{1i},t_{2i})$, and $\mathbf{w}_i=(1,t_{3i})$, where $t_{1i}, t_{2i}$, and $t_{3i}$ were drawn independently from the standard normal model. Note that we considered an intercept in the three linear predictors of the model, with two variables (the same) to model $\mu$ and $\alpha$ and one different variable to model $\beta$. We also considered ${\bm \beta}_1=(-2, 1.5, 0.3)$, ${\bm \beta}_2=(1, -0.4, 0.7)$, and ${\bm \beta}_3=(1, -0.5)$; three sample sizes: $n=100$, $200$, and $305$; and three quantiles to be modeled: $\tau=0.25, 0.50$, and $0.90$. The generation of GB3 random variables was performed based on the stochastic representation in Equation (\ref{re2}).

\subsection{Recovery parameters}
\noindent

In this study, we focused on the properties of the ML estimators of the GB3 regression model. The data generation and parameter estimation were based on the logit link. For each combination of $\tau$ and $n$, we drew 1,000 samples of size $n$ and computed the ML estimates. We present the estimated bias (bias), the mean of the estimated standard errors (SE), the root of the estimated mean squared error (RMSE), and the 95\% coverage probabilities (CP) based on the normal distribution. Table \ref{table.sim.1} summarizes the results. Note that, in general, within each linear predictor, the intercepts presented the greater bias. However, such bias is reduced as the sample size increases. We also note that SE and RMSE decrease when $n$ increases, suggesting that the standard errors are well estimated and the estimators are consistent. Finally, the CP's seem poor for small sample sizes, especially for the parameters related to the quantile. However, this is natural considering that we are estimating eight parameters based on 100 observations. We also remark that such CP's improve considerably for $n=200$ and $n=305$. This result suggests that one should be careful with the asymptotic normal distribution of the ML estimator for small and moderate sample sizes.

\begin{table}[!htbp]
\caption{Empirical bias, SE, RMSE, and CP of the ML estimators of the GB3 model under the indicated values for simulated data.}\label{table.sim.1}
\begin{center}
\resizebox{\linewidth}{!}{
\begin{tabular}{ccrcccrcccrccc}
\hline
& & \multicolumn{4}{c}{$n=100$} & \multicolumn{4}{c}{$n=200$} & \multicolumn{4}{c}{$n=305$} \\
       $\tau$ &  parameter &       bias &         se &       RMSE &         cp &       bias &         se &       RMSE &         cp &       bias &         se &       RMSE &         cp \\
\hline
      0.25 &     $\theta_{0}$ &      0.018 &      0.116 &      0.122 &      0.923 &      0.012 &      0.082 &      0.084 &      0.944 & 0.0076 & 0.0656 & 0.0686 & 0.928\\

           &     $\theta_{1}$ &      0.002 &      0.095 &      0.099 &      0.941 &     -0.003 &      0.072 &      0.073 &      0.946 & 0.0001 & 0.0594 & 0.0596 & 0.957\\

           &     $\theta_{2}$ &     -0.004 &      0.105 &      0.106 &      0.945 &     -0.002 &      0.085 &      0.088 &      0.946 & -0.0001 & 0.0580 & 0.0599 & 0.932\\
\cline{2-14}
           &     $\nu_{0}$ &      0.172 &      0.449 &      0.586 &      0.934 &      0.043 &      0.264 &      0.293 &      0.945 & 0.0372 & 0.2065 & 0.2153 & 0.949\\

           &     $\nu_{1}$ &     0.000 &      0.257 &      0.281 &      0.946 &      0.005 &      0.158 &      0.167 &      0.942 & -0.0055 & 0.1278 & 0.1284 & 0.948\\

           &     $\nu_{2}$ &      0.043 &      0.319 &      0.372 &      0.934 &     -0.006 &      0.144 &      0.147 &      0.949 &  0.0007 & 0.1299 & 0.1399 & 0.932\\
\cline{2-14}

           &     $\eta_{0}$ &      0.245 &      0.438 &      0.702 &      0.958 &      0.110 &      0.270 &      0.323 &      0.962 & 0.0643 & 0.1980 & 0.2265 & 0.964\\

           &     $\eta_{1}$ &     -0.071 &      0.252 &      0.321 &      0.940 &     -0.034 &      0.203 &      0.220 &      0.958 & -0.0202 & 0.1558 & 0.1620 & 0.952\\

\hline
      0.50 &     $\theta_{0}$ &      0.003 &      0.107 &      0.117 &      0.916 &      0.002 &      0.074 &      0.078 &      0.935 & 0.0076 & 0.0656 & 0.0686 & 0.928\\

           &     $\theta_{1}$ &     -0.007 &      0.089 &      0.097 &      0.916 &     -0.005 &      0.067 &      0.067 &      0.945 & 0.0001 & 0.0594 & 0.0596 & 0.957\\

           &     $\theta_{2}$ &      0.005 &      0.096 &      0.103 &      0.931 &      0.002 &      0.078 &      0.079 &      0.946 & -0.0001 & 0.0580 & 0.0599 & 0.932\\
\cline{2-14}

           &     $\nu_{0}$ &      0.216 &      0.492 &      0.635 &      0.954 &      0.083 &      0.287 &      0.325 &      0.957 & 0.0372 & 0.2065 & 0.2153 & 0.949\\

           &     $\nu_{1}$ &     -0.036 &      0.305 &      0.339 &      0.943 &     -0.005 &      0.189 &      0.195 &      0.953 & -0.0055 & 0.1278 & 0.1284 & 0.948\\

           &     $\nu_{2}$ &      0.078 &      0.371 &      0.410 &      0.944 &      0.006 &      0.180 &      0.190 &      0.948 & 0.0007 & 0.1299 & 0.1399 & 0.932\\
\cline{2-14}

           &     $\eta_{0}$ &      0.246 &      0.436 &      0.927 &      0.959 &      0.085 &      0.266 &      0.297 &      0.965 & 0.0643 & 0.1980 & 0.2265 & 0.964\\

           &     $\eta_{1}$ &     -0.057 &      0.249 &      0.311 &      0.947 &     -0.019 &      0.202 &      0.208 &      0.950 & -0.0202 & 0.1558 & 0.1620 & 0.952\\

\hline
      0.90 &    $\theta_{0}$ &     -0.035 &      0.135 &      0.144 &      0.921 &     -0.013 &      0.096 &      0.098 &      0.942 & 0.0076 & 0.0656 & 0.0686 & 0.928\\

           &     $\theta_{1}$ &     -0.011 &      0.101 &      0.106 &      0.931 &     -0.002 &      0.076 &      0.076 &      0.953 & 0.0001 & 0.0594 & 0.0596 & 0.957\\

           &     $\theta_{2}$ &      0.012 &      0.109 &      0.116 &      0.930 &      0.006 &      0.085 &      0.089 &      0.936 & -0.0001 & 0.0580 & 0.0599 & 0.932\\
\cline{2-14}

           &     $\nu_{0}$ &      0.283 &      0.503 &      0.678 &      0.959 &      0.110 &      0.291 &      0.332 &      0.960 & 0.0372 & 0.2065 & 0.2153 & 0.949\\

           &     $\nu_{1}$ &     -0.033 &      0.317 &      0.362 &      0.946 &     -0.022 &      0.191 &      0.199 &      0.951 & -0.0055 & 0.1278 & 0.1284 & 0.948\\

           &     $\nu_{2}$ &      0.090 &      0.382 &      0.417 &      0.947 &      0.020 &      0.189 &      0.188 &      0.961 & 0.0007 & 0.1299 & 0.1399 & 0.932\\
\cline{2-14}

           &     $\eta_{0}$ &      0.109 &      0.357 &      0.433 &      0.956 &      0.044 &      0.242 &      0.279 &      0.961 & 0.0643 & 0.1980 & 0.2265 & 0.964\\

           &     $\eta_{1}$ &     -0.022 &      0.152 &      0.169 &      0.930 &     -0.005 &      0.134 &      0.147 &      0.931 & -0.0202 & 0.1558 & 0.1620 & 0.952\\
\hline

\end{tabular}
}
\end{center}
\end{table}

\subsection{Assessing the choice of the links}
\noindent

In this section, we show that, given the ``true link'' used for the quantile, the performance of the ML estimators of the GB3 regression model is good. However, the correct selection of the link can be a critical point in a real data application. We considered only the estimated log-likelihood (LL) function for model selection and the logit, probit, loglog, and cloglog link functions. The data generation was performed based on the GB3 model and a specified ``true link'' (from among those mentioned), and the estimation based on the GB3 model and the four links. For each case, we computed the estimated LL function, the root of the mean square prediction error (MSPE), and the root of the mean absolute prediction error (MAPE), in both cases for the quantiles. Such measures are given by
\[
\mbox{MSPE}=\sum_{i=1}^n (\mu_i-\widehat{\mu}_i)^2/n \qquad \mbox{and} \qquad \mbox{MAPE}=\sum_{i=1}^n |\mu_i-\widehat{\mu}_i|/n,
\]
where $\mu_i$ and $\widehat{\mu}_i$ denote the true and estimated quantile for the $i$-th observation, respectively.
In other words, we focused on the case where the estimation of the quantile is more important than the shapes of the distribution.
For this reason, the MSPE and MAPE represent the ``true'' criteria to select a link based on two different distance measures between the true and estimated quantile. We also highlight that a greater LL suggests a better fit, whereas a lower MSPE and MAPE suggest the model is well adjusted.
Provided the different links do not modify the numbers of parameters, the use of the LL as a model selection criterion is equivalent to the use of AIC and BIC. Tables \ref{table.sim.2} and \ref{table.sim.2.2} report the percentage of times where each criterion (LL, MSPE, and MAPE) selected the links for each combination of true link, $\tau$, and $n$ for 1,000 replicates. Overall, the model selection based on LL (AIC or BIC) seems a reasonable criterion to select a link if the interest lies in the estimation of the quantile.

\begin{table}[!htbp]
\caption{Empirical percentage of times in which LL, MSPE, and MAPE selected the logit and probit link functions for the GB3 regression model.}\label{table.sim.2}
\begin{center}
\resizebox{\linewidth}{!}{
\begin{tabular}{cccrrrrrrrrr}
\hline
      true &            &        fitted    &            \multicolumn{ 3}{c}{$n=100$} &           \multicolumn{ 3}{c}{$n=200$} &           \multicolumn{ 3}{c}{$n=305$} \\
      link &        tau &       link &         LL &       MSPE &       MAPE &         LL &       MSPE &       MAPE &         LL &       MSPE &       MAPE \\
\hline

     logit &       0.25 &      logit & {\bf 55.3} & {\bf 71.3} & {\bf 61.0} & {\bf 82.2} & {\bf 87.7} & {\bf 80.4} &  {\bf 93.1} &  {\bf 96.9} &   {\bf 89.5}\\

           &            &     probit &  19.2 &       15.8 &       13.8 &       12.1 &        7.6 &        3.7 & 4.3 &   3.0 &   0.9 \\

           &            &     loglog &  0.8 &        0.0 &        0.1 &        0.0 &        0.0 &        0.0 & 0.0  &  0.0  &  0.0\\

           &            &    cloglog &  24.7 &       12.9 &       25.1 &        5.7 &        4.7 &       15.9 & 2.6 &   0.1  &  9.6\\
\cline{2-12}
           &        0.5 &      logit & {\bf 52.4} & {\bf 75.3} & {\bf 70.5} & {\bf 82.8} & {\bf 88.3} & {\bf 82.6} & {\bf 93.4} &  {\bf 97.5} &  {\bf 91.9} \\

           &            &     probit &  18.1 &       11.2 &       10.3 &       10.5 &        9.8 &        3.0 & 4.1 &   2.5 &   0.5\\

           &            &     loglog &  0.7 &        0.0 &        0.0 &        0.0 &        0.0 &        0.0 &  0.0 &   0.0  &  0.0\\

           &            &    cloglog &  28.8 &       13.5 &       19.2 &        6.7 &        1.9 &       14.4 & 2.5 &   0.0 &   7.6\\
\cline{2-12}

           &        0.9 &      logit & {\bf 57.7} & {\bf 72.4} & {\bf 72.6} & {\bf 84.2} & {\bf 90.1} & {\bf 84.5} & {\bf 93.3} &  {\bf 96.2} &  {\bf 90.9} \\

           &            &     probit &  15.3 &       10.5 &        7.9 &       10.0 &        7.9 &        3.0 & 2.9  &  3.4  &  0.3\\

           &            &     loglog &   0.5 &        0.0 &        0.0 &        0.0 &        0.0 &        0.0 &  0.0 &   0.0  &  0.0\\

           &            &    cloglog &  26.5 &       17.1 &       19.5 &        5.8 &        2.0 &       12.5 & 3.8 &   0.4  &  8.8\\
\hline
    probit &       0.25 &      logit &   0.0 &        0.0 &        0.0 &        0.0 &        0.0 &        0.1 & 0.0 &   0.0  &  0.0\\

           &            &     probit & {\bf 99.9} & {\bf 100.0} & {\bf 100.0} & {\bf 100.0} & {\bf 100.0} & {\bf 99.9} & {\bf 100.0} &  {\bf 100.0} & {\bf 100.0} \\

           &            &     loglog &  0.1 &        0.0 &        0.0 &        0.0 &        0.0 &        0.0 & 0.0 &   0.0  &  0.0\\

           &            &    cloglog &   0.0 &        0.0 &        0.0 &        0.0 &        0.0 &        0.0 & 0.0 &   0.0 &   0.0\\
\cline{2-12}

           &        0.5 &      logit &     0.0 &        0.2 &        0.3 &        0.0 &        0.0 &        0.0 & 0.0 &   0.0 &   0.0\\

           &            &     probit & {\bf 100.0} & {\bf 99.8} & {\bf 99.4} & {\bf 100.0} & {\bf 100.0} & {\bf 100.0} & {\bf 100.0} &  {\bf 100.0} & {\bf 100.0} \\

           &            &     loglog &    0.0 &        0.0 &        0.0 &        0.0 &        0.0 &        0.0 & 0.0 &   0.0 &   0.0\\

           &            &    cloglog &    0.0 &        0.0 &        0.3 &        0.0 &        0.0 &        0.0 & 0.0  &  0.0 &   0.0\\
\cline{2-12}

           &        0.9 &      logit &    0.0 &        2.0 &        2.8 &        0.0 &        0.3 &        0.8 & 0.0 &   0.0  &  0.1\\

           &            &     probit & {\bf 100.0} & {\bf 98.0} & {\bf 96.3} & {\bf 100.0} & {\bf 99.7} & {\bf 99.2} & {\bf 100.0} & {\bf 100.0}  & {\bf 99.9} \\

           &            &     loglog &    0.0 &        0.0 &        0.0 &        0.0 &        0.0 &        0.0 & 0.0  &  0.0  &  0.0\\

           &            &    cloglog &    0.0 &        0.0 &        0.9 &        0.0 &        0.0 &        0.0 & 0.0  &  0.0  &  0.0\\

\hline
\end{tabular}
}
\end{center}
\end{table}

\begin{table}[!htbp]
\caption{Empirical percentage of times in which LL, MSPE, and MAPE selected the loglog and cloglog link functions for the GB3 regression model.}\label{table.sim.2.2}
\begin{center}
\resizebox{\linewidth}{!}{
\begin{tabular}{cccrrrrrrrrr}
\hline
      true &            &        fitted    &            \multicolumn{ 3}{c}{$n=100$} &           \multicolumn{ 3}{c}{$n=200$} &           \multicolumn{ 3}{c}{$n=305$} \\
      link &        tau &       link &         LL &       MSPE &       MAPE &         LL &       MSPE &       MAPE &         LL &       MSPE &       MAPE \\
\hline
    loglog &       0.25 &      logit &        0.0 &        0.0 &        0.0 &        0.0 &        0.0 &        0.0 & 0.0 &   0.0 &   0.4 \\

           &            &     probit &        0.0 &        0.0 &        0.0 &        0.0 &        0.0 &        0.0 & 0.0 &   0.0 &   0.0\\

           &            &     loglog & {\bf 100.0} & {\bf 100.0} & {\bf 100.0} & {\bf 100.0} & {\bf 100.0} & {\bf 100.0} & {\bf 100.0} &  {\bf 100.0} &  {\bf 99.6}\\

           &            &    cloglog &        0.0 &        0.0 &        0.0 &        0.0 &        0.0 &        0.0 & 0.0 &   0.0 &   0.0 \\
\cline{2-12}

           &        0.5 &      logit &        0.0 &        0.0 &        0.0 &        0.0 &        0.0 &        0.0 & 0.1 &   0.0 &   0.2 \\

           &            &     probit &        0.0 &        0.0 &        0.0 &        0.0 &        0.0 &        0.0 & 0.0 &   0.0  &  0.0\\

           &            &     loglog & {\bf 100.0} & {\bf 100.0} & {\bf 100.0} & {\bf 100.0} & {\bf 100.0} & {\bf 100.0} & {\bf 99.9} & {\bf 100.0} &   {\bf 99.8} \\

           &            &    cloglog &        0.0 &        0.0 &        0.0 &        0.0 &        0.0 &        0.0 & 0.0  &  0.0  &  0.0\\
\cline{2-12}

           &        0.9 &      logit &        0.1 &        0.1 &        0.1 &        0.0 &        0.0 &        0.0 & 0.8  &  0.1 &   0.2\\

           &            &     probit &        0.0 &        0.0 &        0.0 &        0.0 &        0.0 &        0.0 & 0.0 &   0.0 &   0.0\\

           &            &     loglog & {\bf 99.9} & {\bf 99.9} & {\bf 99.9} & {\bf 100.0} & {\bf 100.0} & {\bf 100.0} & {\bf 99.2} &  {\bf 99.9}  & {\bf 99.8} \\

           &            &    cloglog &        0.0 &        0.0 &        0.0 &        0.0 &        0.0 &        0.0 & 0.0 &   0.0  &  0.0\\
\hline
   cloglog &       0.25 &      logit &       19.1 &        4.7 &        9.7 &        1.6 &        0.4 &        3.8 & 0.0 &   0.1 &   1.0\\

           &            &     probit &        4.1 &        0.1 &        0.2 &        0.1 &        0.0 &        0.0 & 0.0 &   0.0  &  0.0\\

           &            &     loglog &        0.3 &        0.0 &        0.0 &        0.0 &        0.0 &        0.0 & 0.0 &   0.0  &  0.0\\

           &            &    cloglog & {\bf 76.5} & {\bf 95.2} & {\bf 90.1} & {\bf 98.3} & {\bf 99.6} & {\bf 96.2} & {\bf 100.0} &  {\bf 99.9} &  {\bf 99.0} \\
\cline{2-12}

           &        0.5 &      logit &       18.4 &        1.4 &        5.9 &        0.1 &        0.4 &        4.6 & 0.0  &  0.0 &   0.6\\

           &            &     probit &        0.6 &        0.0 &        0.0 &        0.0 &        0.0 &        0.0 & 0.0 &   0.0  &  0.0\\

           &            &     loglog &        0.0 &        0.0 &        0.0 &        0.0 &        0.0 &        0.0 & 0.0 &   0.0  &  0.0\\

           &            &    cloglog & {\bf 81.0} & {\bf 98.6} & {\bf 94.1} & {\bf 99.9} & {\bf 99.6} & {\bf 95.4} & {\bf 100.0} & {\bf 100.0} &  {\bf 99.4} \\
\cline{2-12}

           &        0.9 &      logit &       16.8 &        1.8 &        9.0 &        0.1 &        0.4 &        2.3 & 0.0 &   0.0 &   1.6\\

           &            &     probit &        0.9 &        0.0 &        0.0 &        0.0 &        0.0 &        0.0 & 0.0  &  0.0  &  0.0\\

           &            &     loglog &        0.0 &        0.0 &        0.0 &        0.0 &        0.0 &        0.0 & 0.0  &  0.0 &   0.0\\

           &            &    cloglog & {\bf 82.3} & {\bf 98.2} & {\bf 91.0} & {\bf 99.9} & {\bf 99.6} & {\bf 97.7} & {\bf 100.0} & {\bf 100.0} &  {\bf 98.4}\\

\hline

\end{tabular}
}
\end{center}
\end{table}

\section{Application to Chilean COVID-19 data}\label{sec:5}
\noindent

In this section, we present and discuss a real data application related to the CFR of COVID-19 in different communes of Chile to illustrate the performance of the GB3 quantile regression model. In order to estimate the parameters of the model, we adopted the ML method (as discussed in Subsection \ref{ML}) and all computations were performed using the function \texttt{optim}($\cdot$) in R.

\subsection{COVID-19 dataset in Chile}
\noindent

\begin{figure}[!htbp]
\vspace{-0.30cm}
\centering
\begin{minipage}[b]{0.46\linewidth}
\centering
\includegraphics[width=7cm]{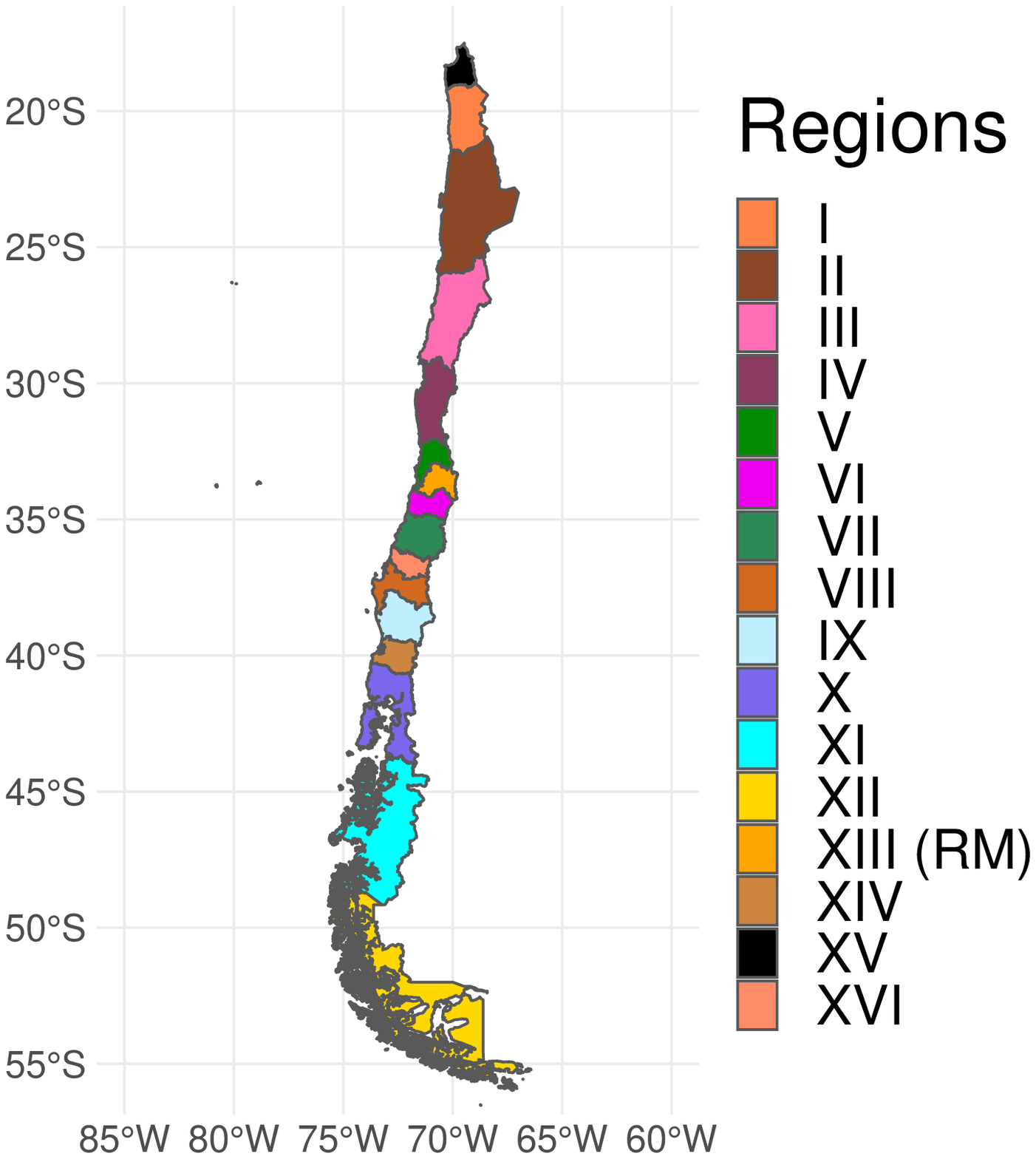}
\end{minipage} 
\hspace{0.2cm}
\begin{minipage}[b]{0.46\linewidth}
\centering
\includegraphics[width=7cm]{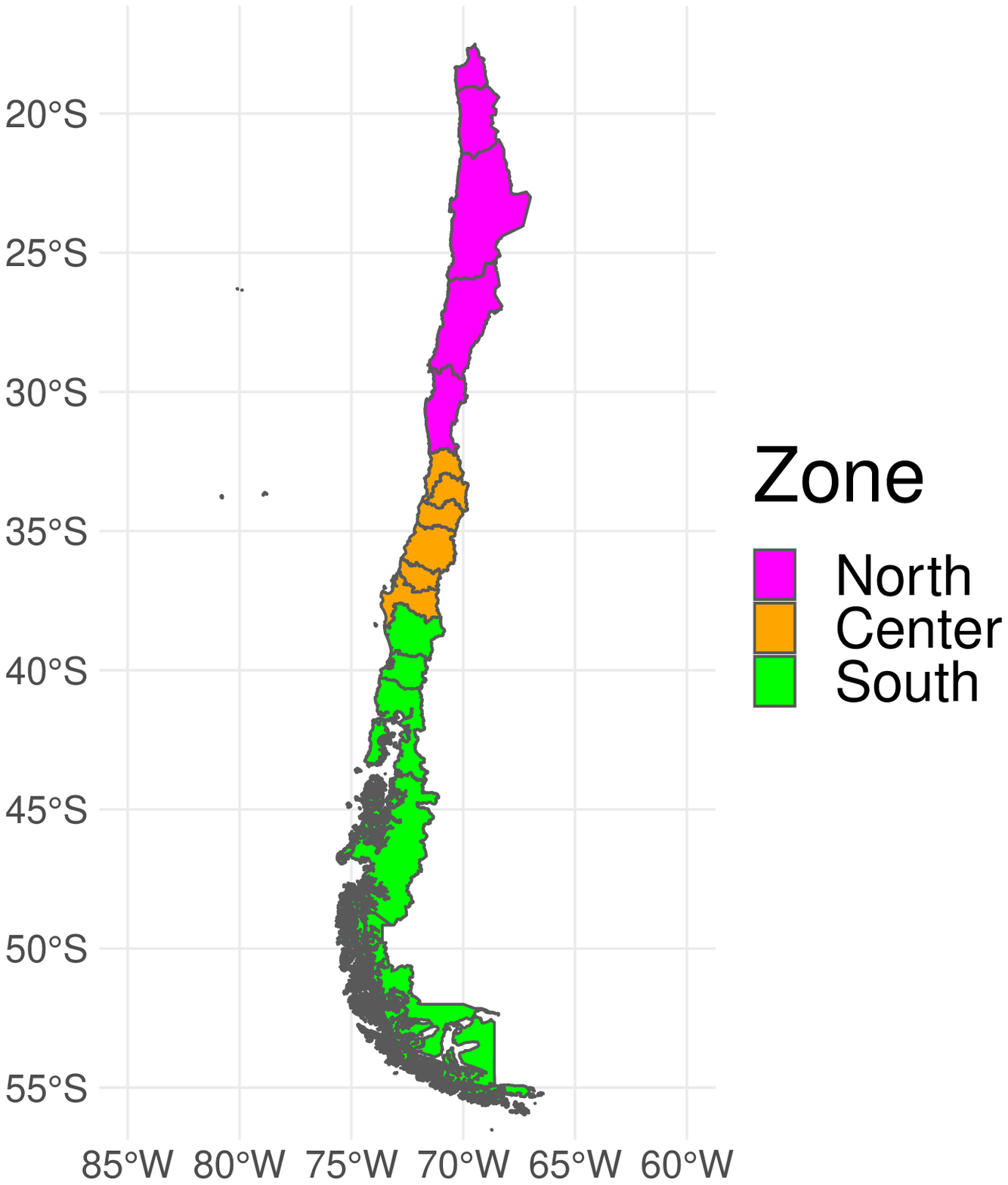}
\end{minipage}\\
\caption{Map of Chile regions and the three different zones.}\label{figchile}
\end{figure}
\begin{figure}[!htbp]
\begin{center}
\resizebox{\linewidth}{!}{
\begin{tabular}{cccc}
North & Center & South \\
  \begin{minipage}{.3\textwidth}{\includegraphics[width=1.3\textwidth]{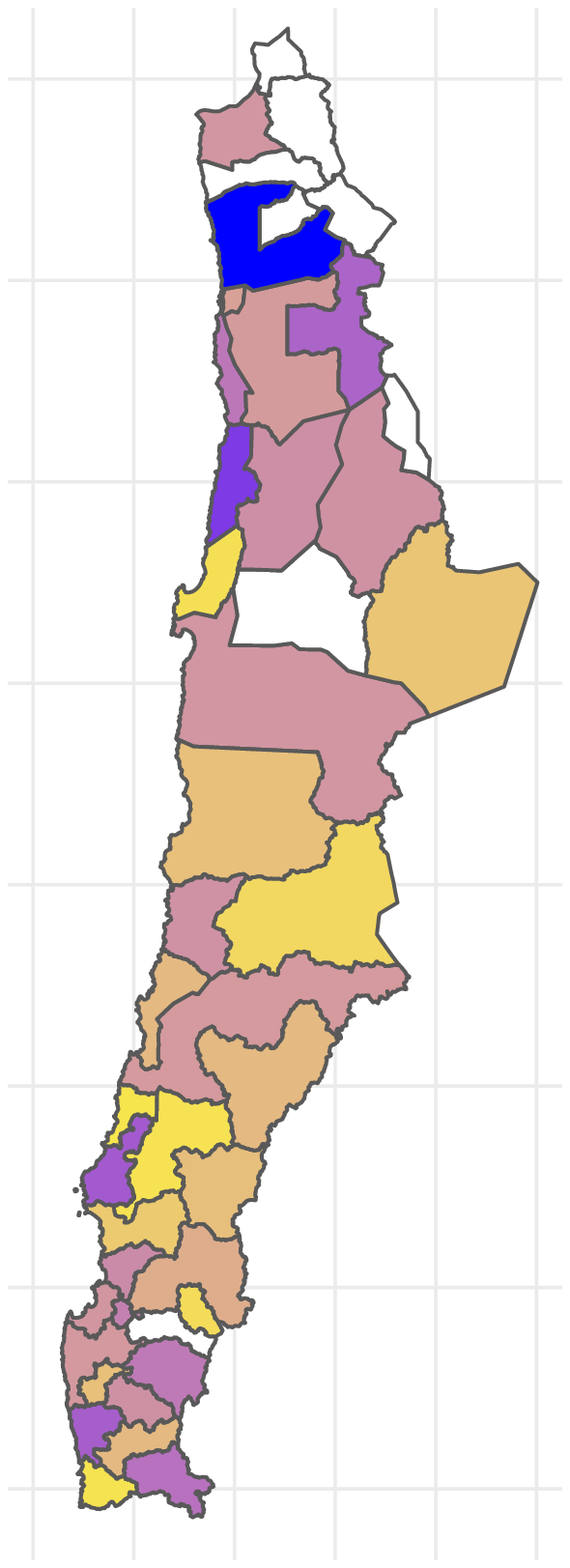}} \end{minipage} &
  \begin{minipage}{.3\textwidth}{\includegraphics[width=1.3\textwidth]{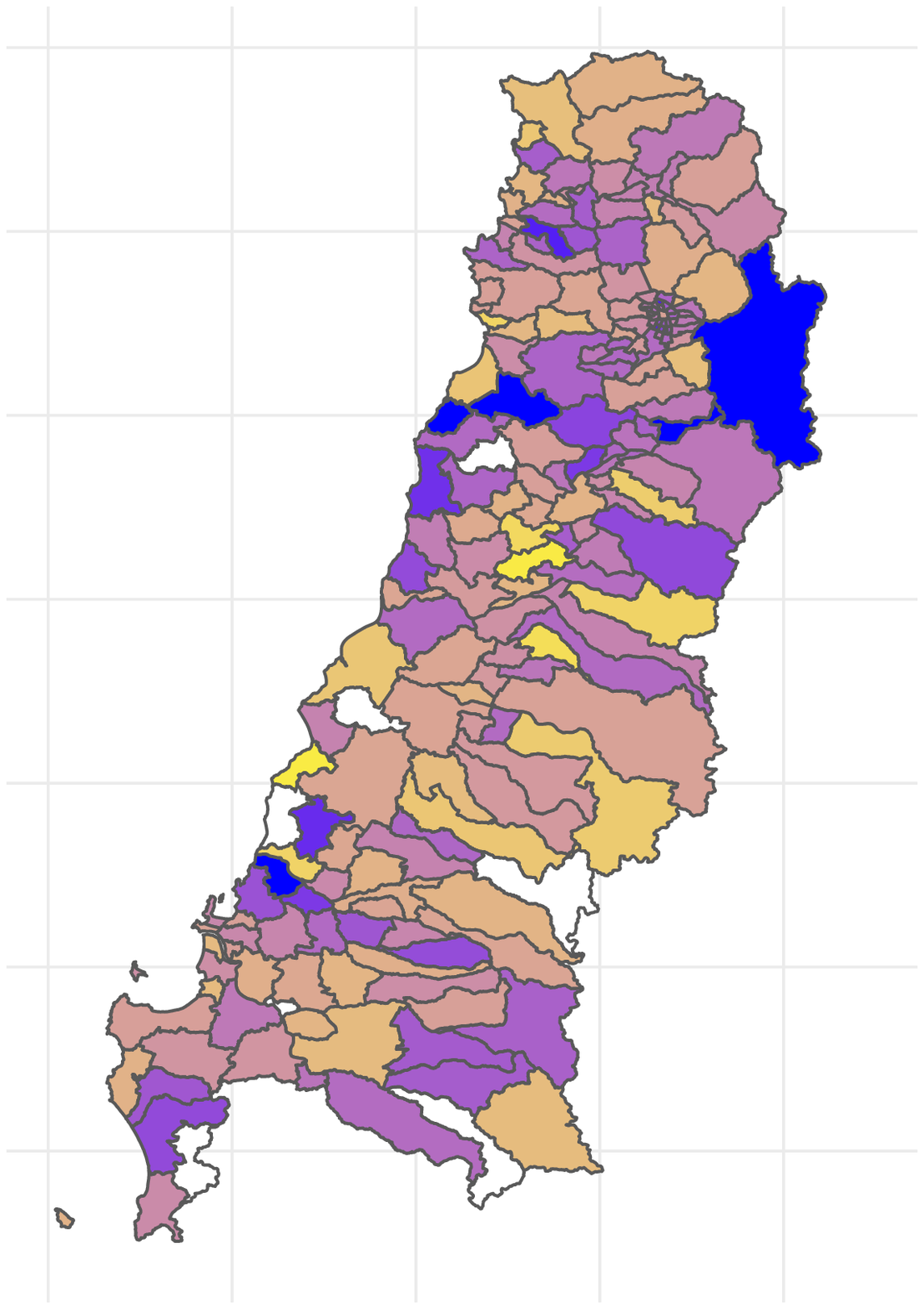}} \end{minipage} &
  \begin{minipage}{.3\textwidth}{\includegraphics[width=1.3\textwidth]{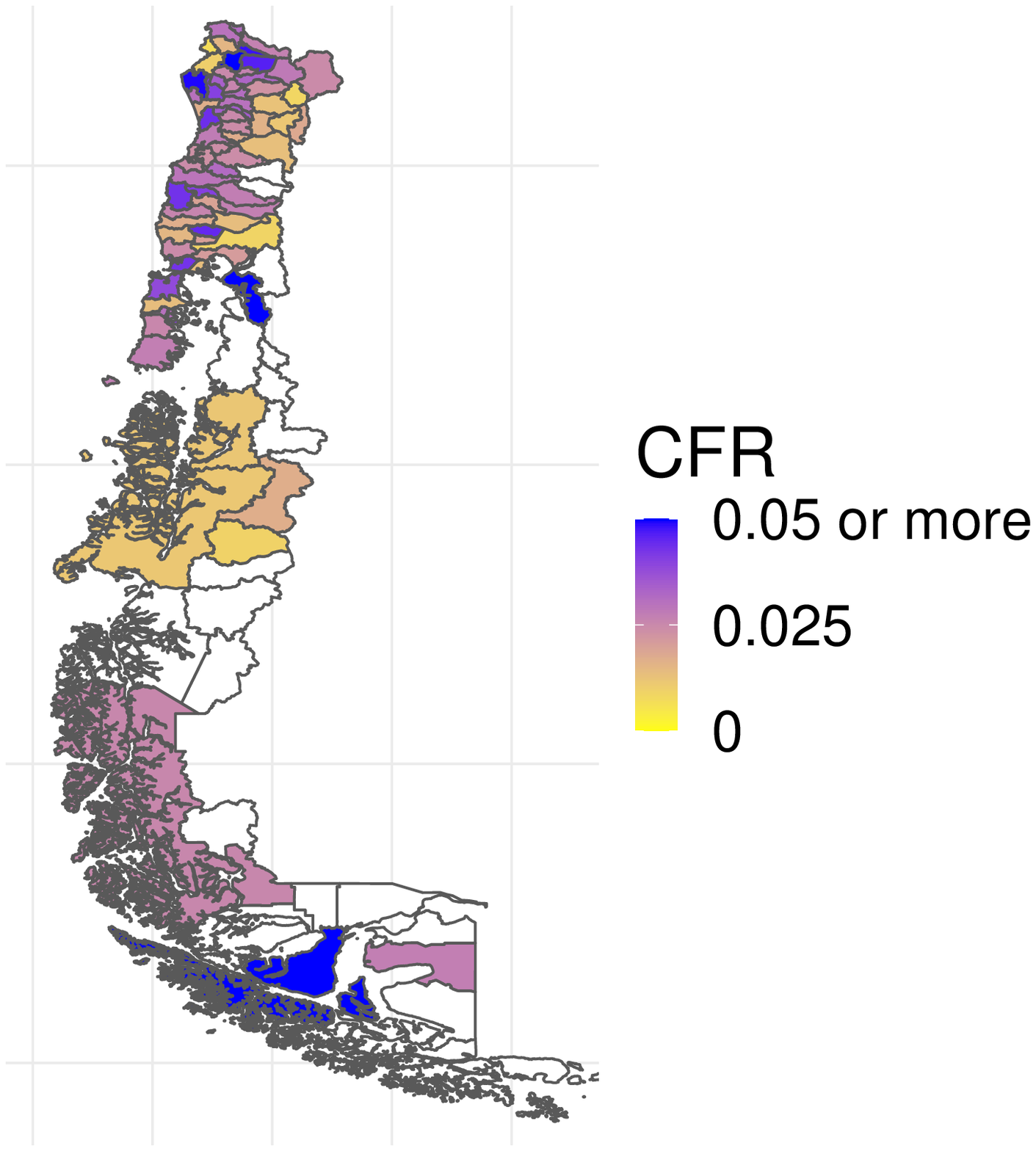}} \end{minipage} \\
\end{tabular}
}
\end{center}
\caption{Averages of the observed CFR values in communes of the North, Center, and South zones of Chile between May 24 and July 23, 2021.}
\label{figchile2}
\end{figure}
\begin{figure}[!htbp]
\begin{center}
\resizebox{\linewidth}{!}{
\begin{tabular}{cccc}
    & North & Center & South \\
  Pos. rate &
  \begin{minipage}{.28\textwidth}{\includegraphics[width=1.2\textwidth]{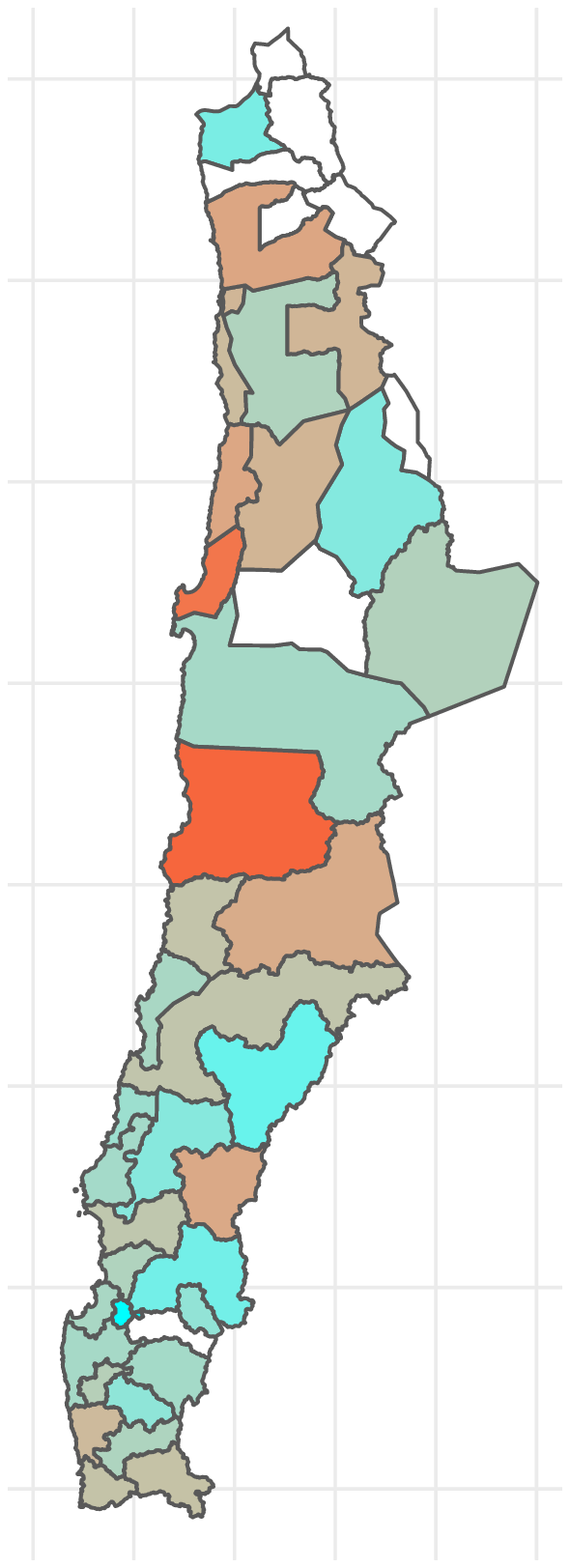}} \end{minipage} &
  \begin{minipage}{.28\textwidth}{\includegraphics[width=1.2\textwidth]{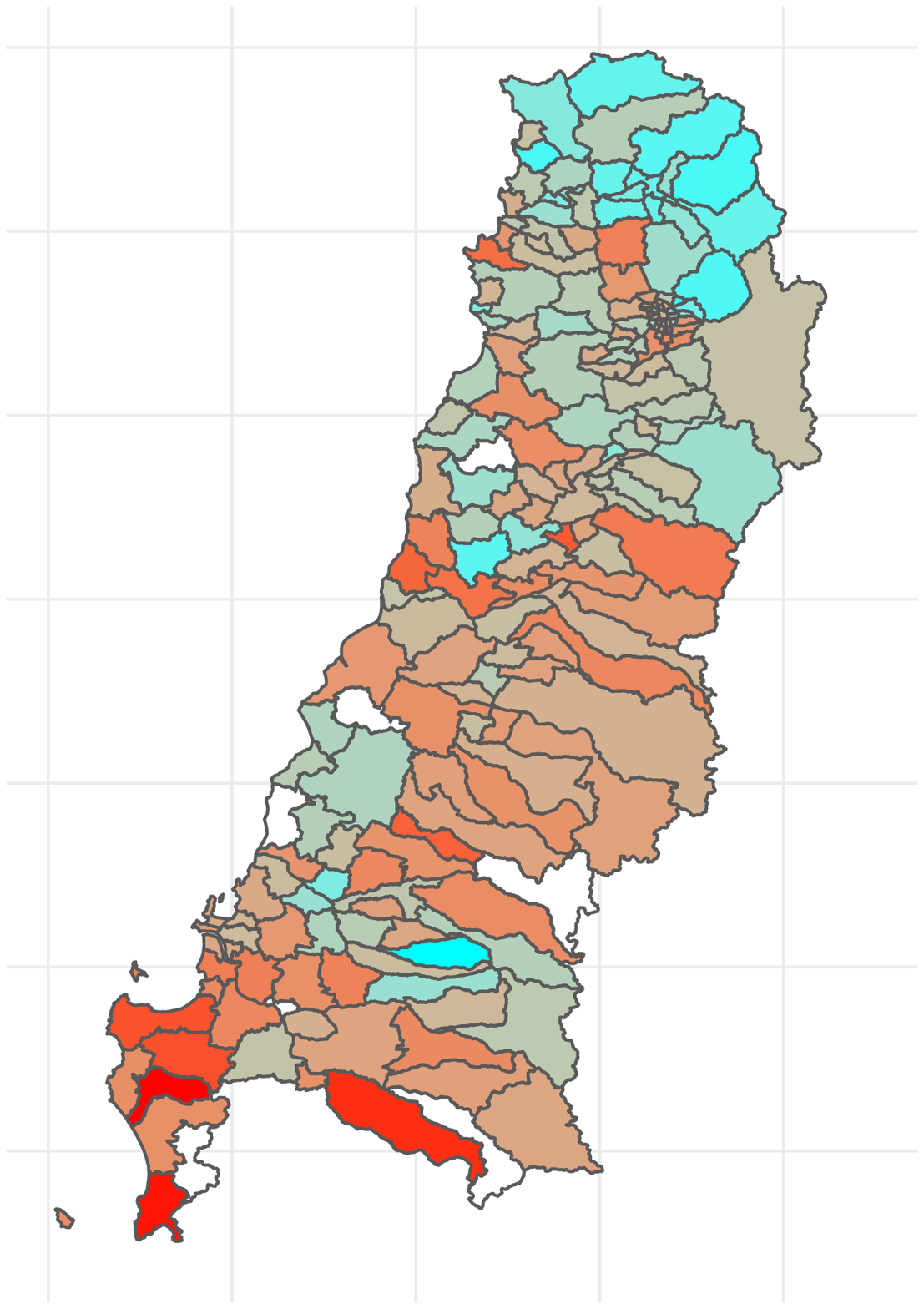}} \end{minipage} &
  \begin{minipage}{.28\textwidth}{\includegraphics[width=1.2\textwidth]{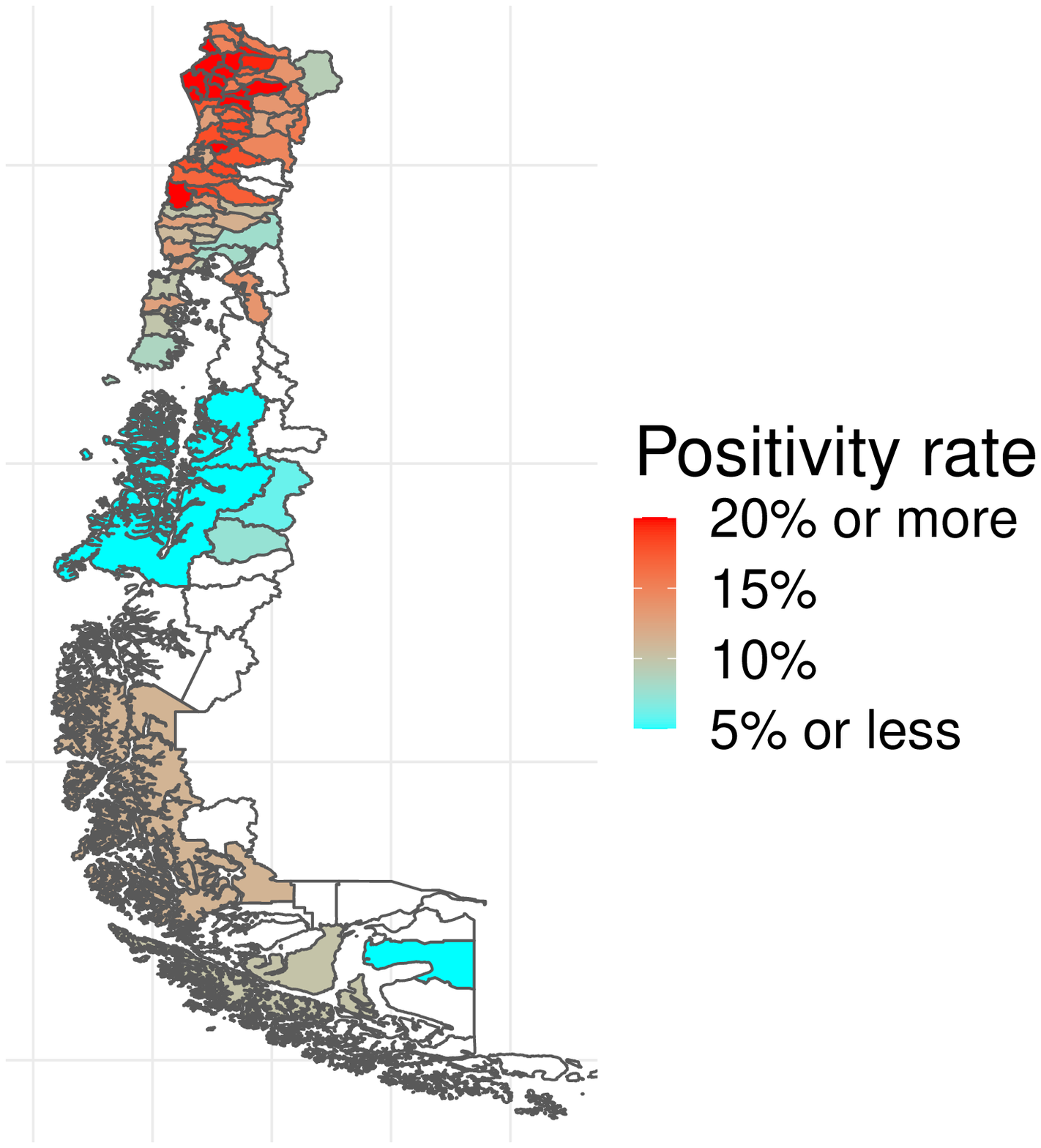}} \end{minipage} \\
  Pop. dens. &
  \begin{minipage}{.28\textwidth}{\includegraphics[width=1.2\textwidth]{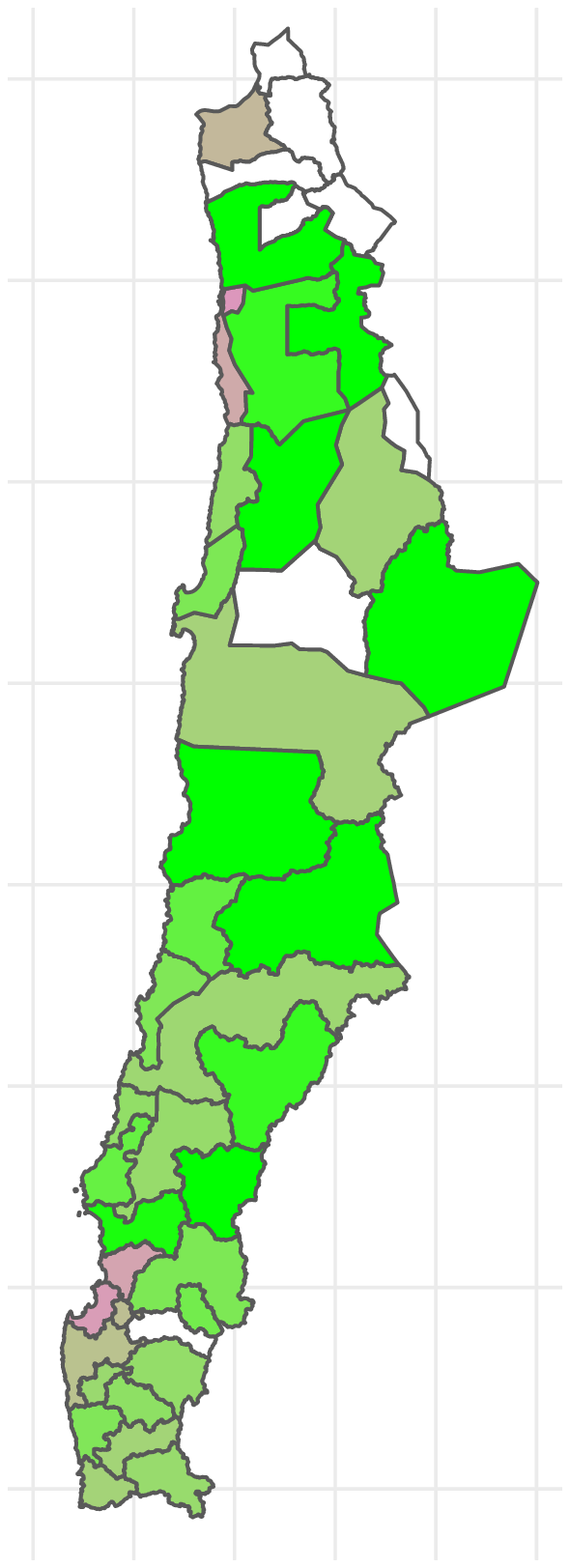}} \end{minipage} &
  \begin{minipage}{.28\textwidth}{\includegraphics[width=1.2\textwidth]{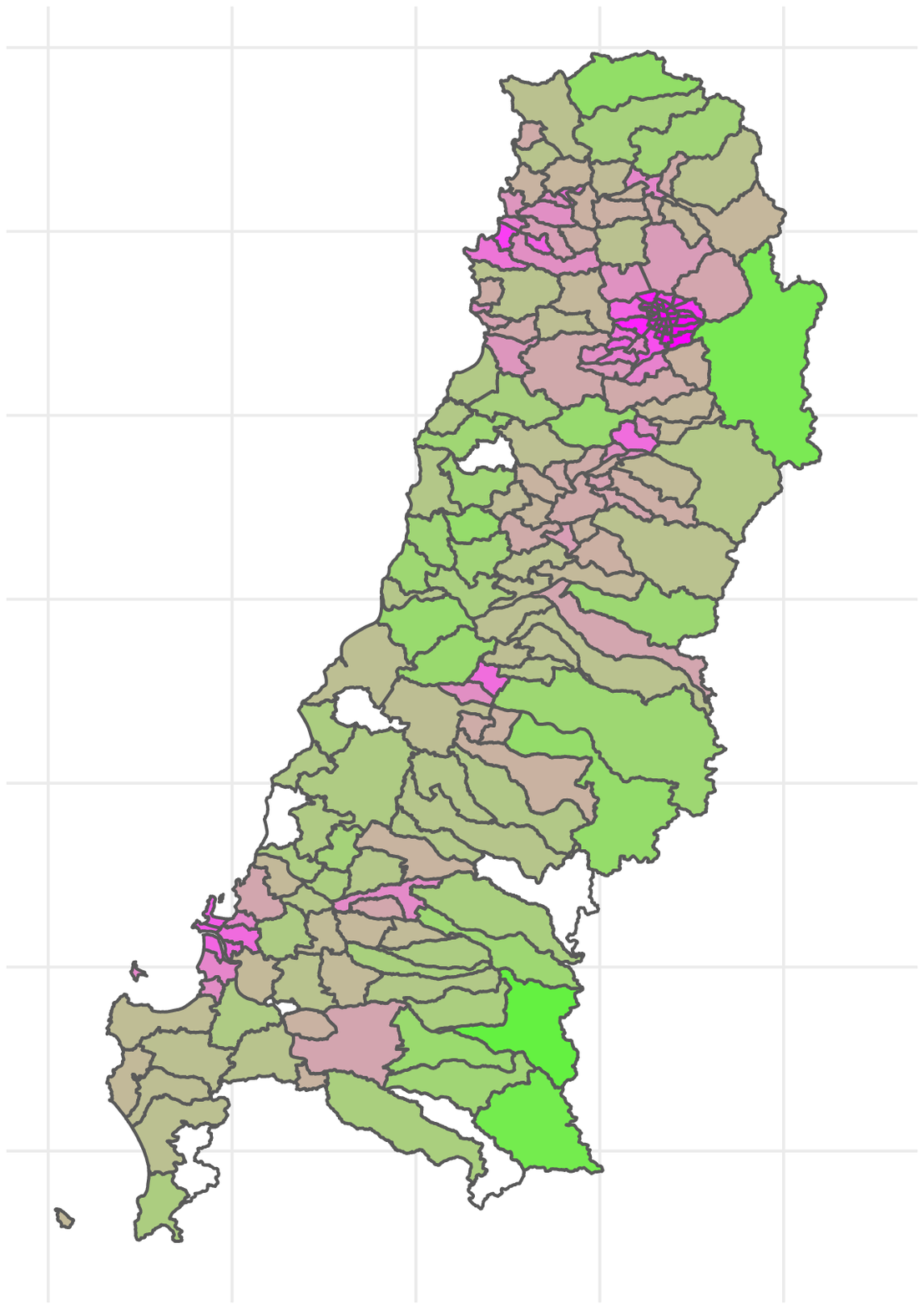}} \end{minipage} &
  \begin{minipage}{.28\textwidth}{\includegraphics[width=1.2\textwidth]{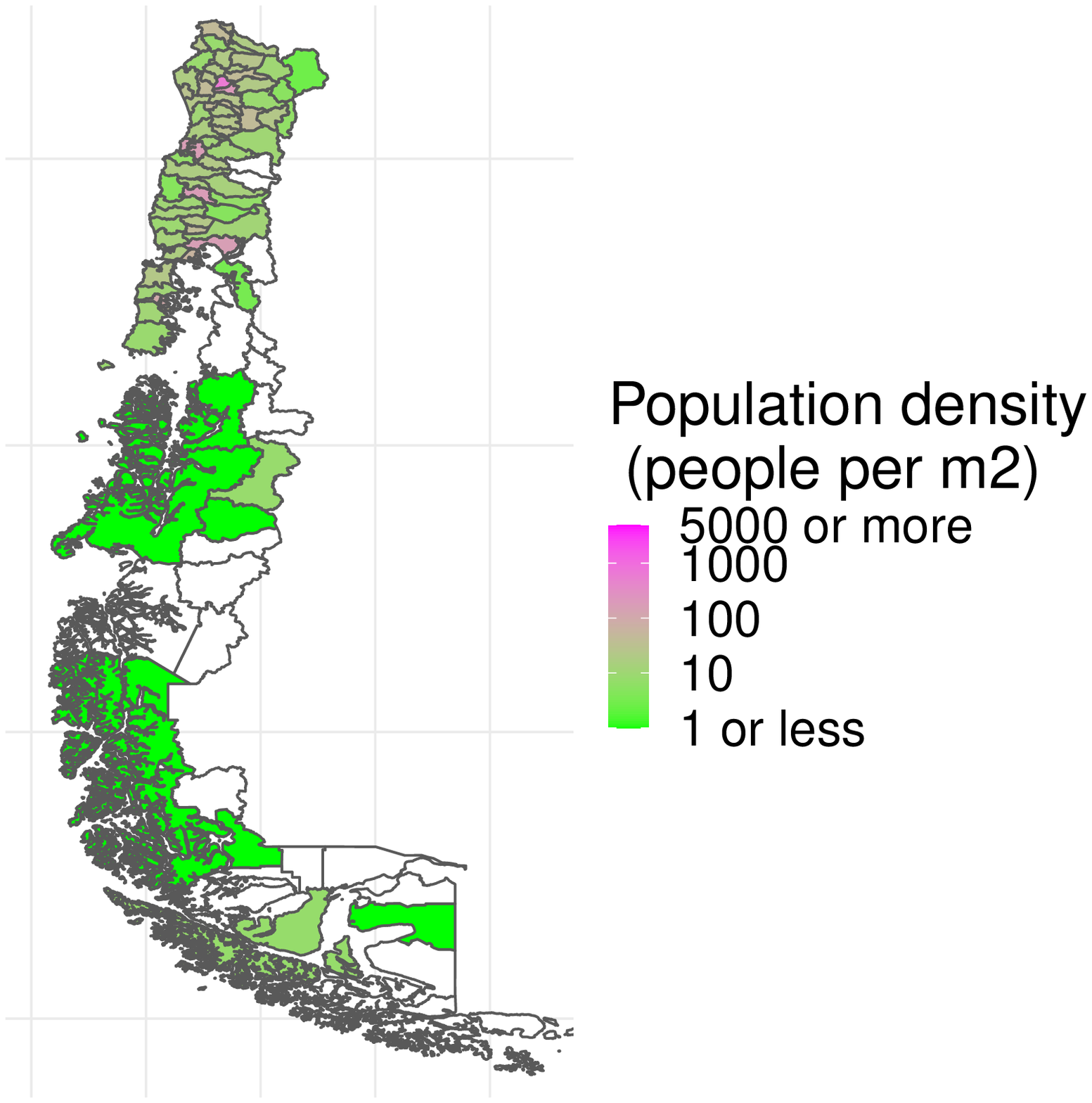}} \end{minipage} \\
  Vacc. &
  \begin{minipage}{.28\textwidth}{\includegraphics[width=1.2\textwidth]{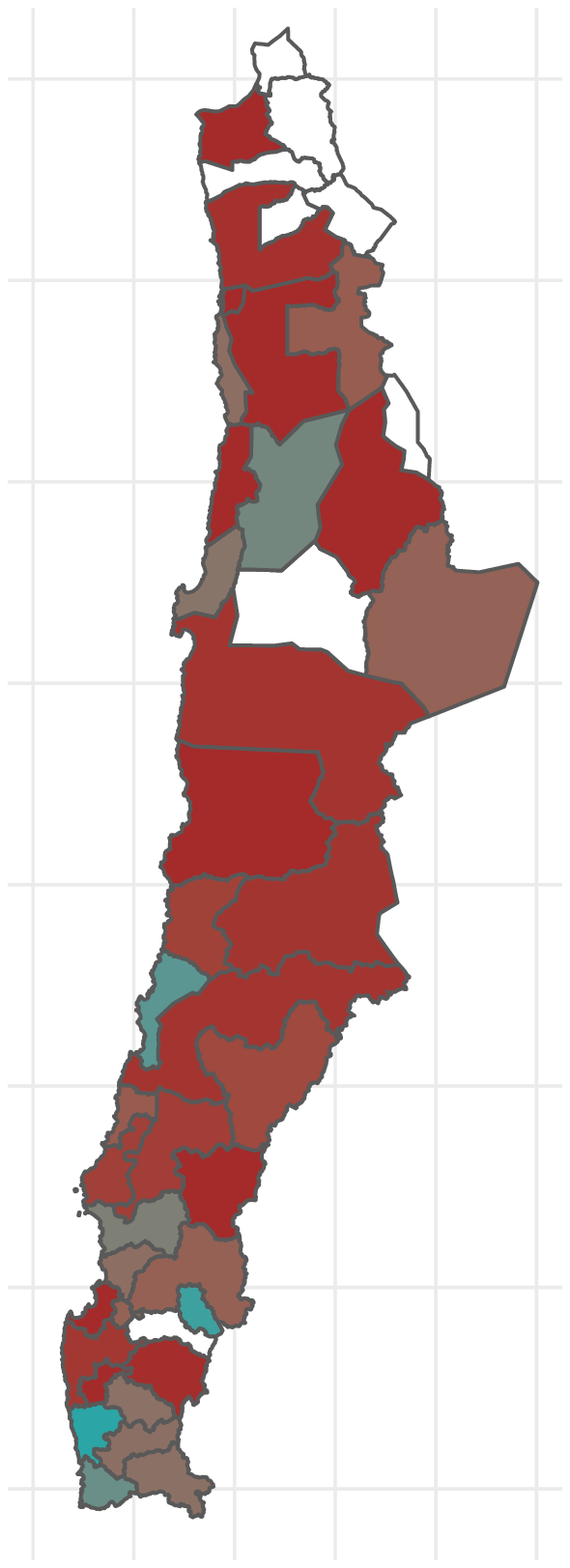}} \end{minipage} &
  \begin{minipage}{.28\textwidth}{\includegraphics[width=1.2\textwidth]{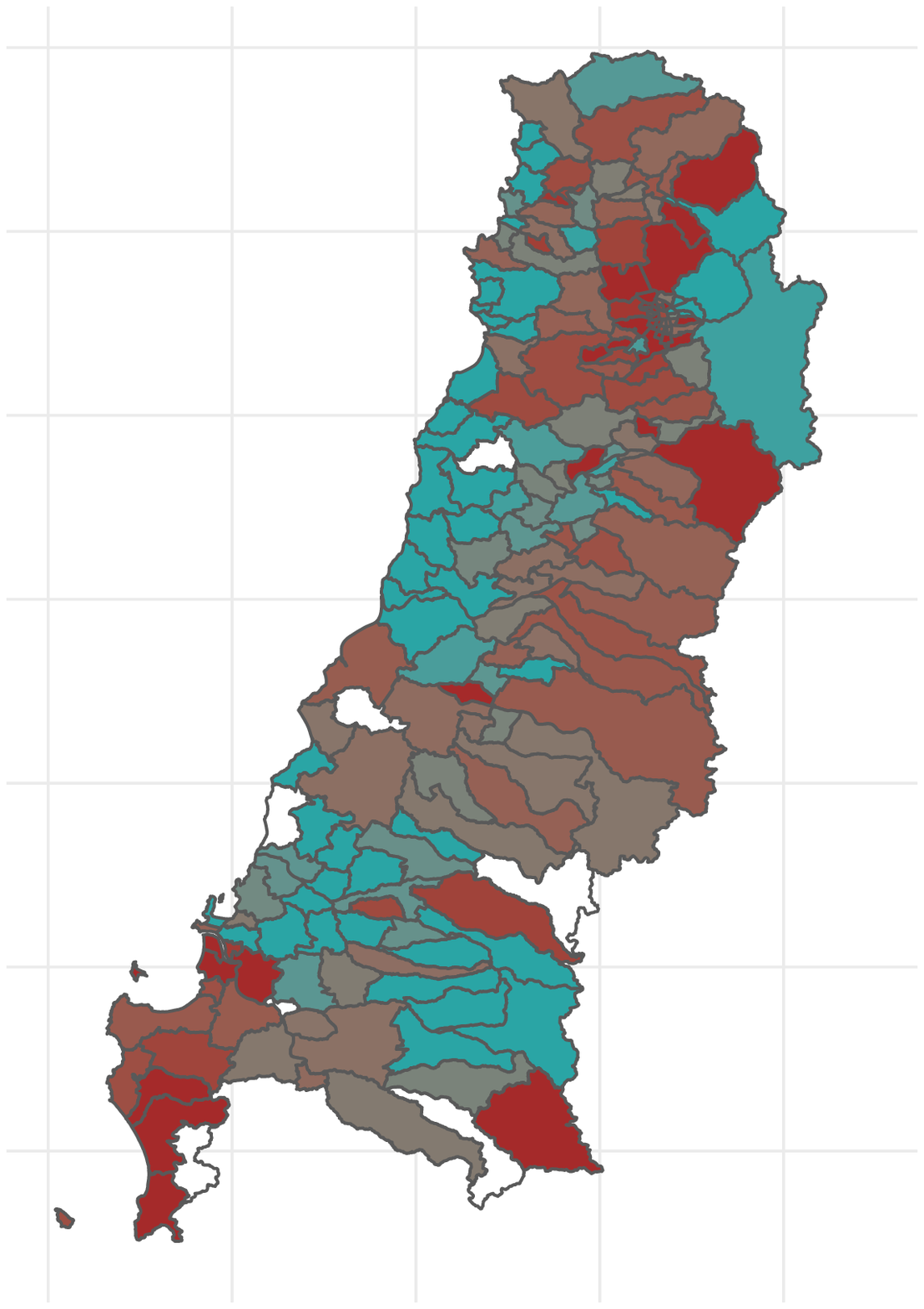}} \end{minipage} &
  \begin{minipage}{.28\textwidth}{\includegraphics[width=1.2\textwidth]{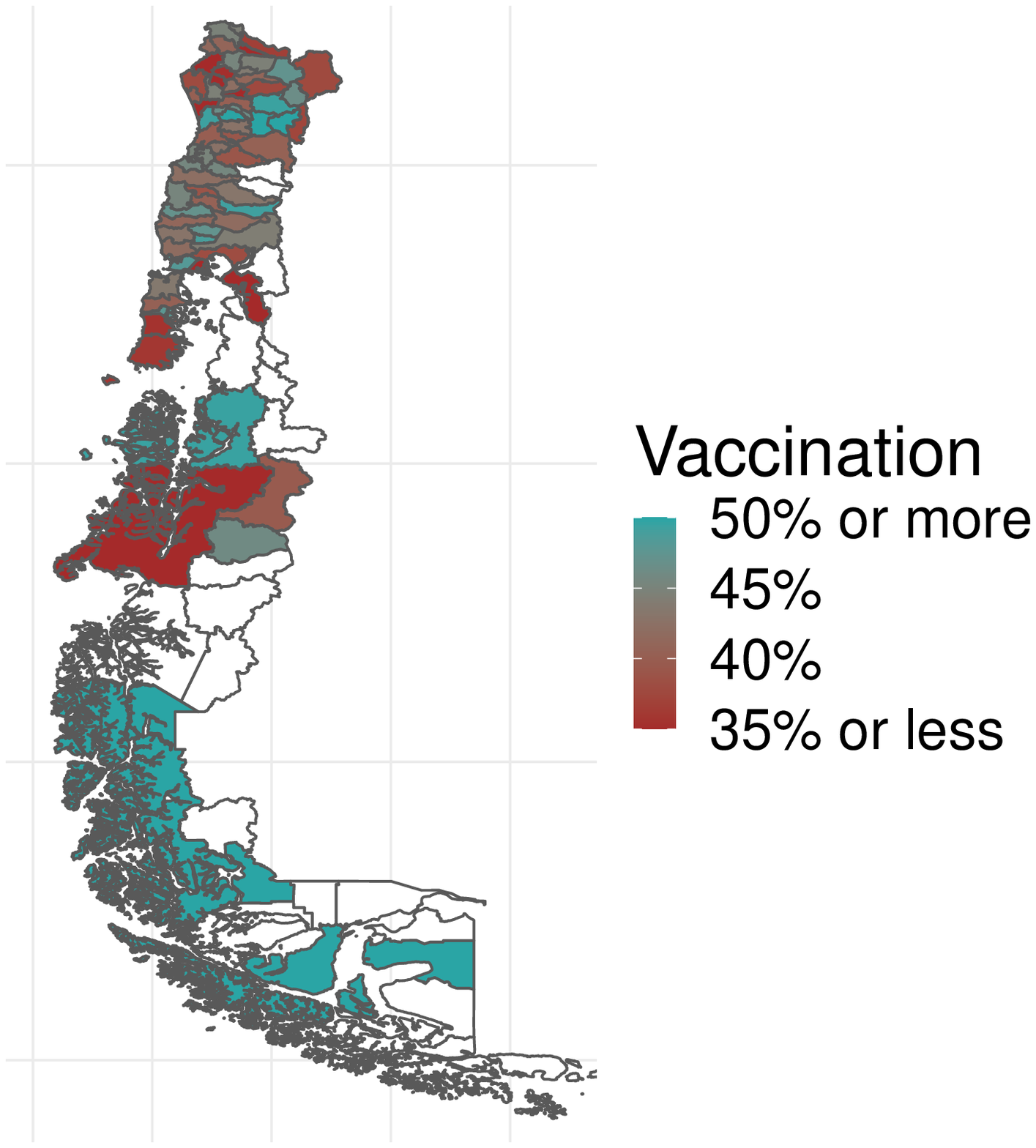}} \end{minipage} \\
\end{tabular}
}
\end{center}
\caption{Positivity rate, population density, and vaccination for communes in the North, Center, and South zones of Chile.}\label{figchile3}
\end{figure}
\begin{figure}[!htbp]
\psfrag{0.514}[c][c]{\tiny{0.514}}
\psfrag{0.512}[c][c]{\tiny{0.512}}
\psfrag{0.510}[c][c]{\tiny{}}
\psfrag{0.508}[c][c]{\tiny{}}
\psfrag{0.506}[c][c]{\tiny{}}
\psfrag{0.504}[c][c]{\tiny{}}
\psfrag{0.502}[c][c]{\tiny{0.502}}
\psfrag{0.500}[c][c]{\tiny{0.500}}
\psfrag{0.520}[c][c]{\tiny{0.520}}
\psfrag{0.515}[c][c]{\tiny{}}
\psfrag{0.505}[c][c]{\tiny{}}
\psfrag{0.385}[c][c]{\tiny{0.385}}
\psfrag{0.380}[c][c]{\tiny{}}
\psfrag{0.375}[c][c]{\tiny{}}
\psfrag{0.370}[c][c]{\tiny{0.370}}
\psfrag{0.650}[c][c]{\tiny{0.650}}
\psfrag{0.645}[c][c]{\tiny{}}
\psfrag{0.640}[c][c]{\tiny{}}
\psfrag{0.635}[c][c]{\tiny{0.635}}
\psfrag{BxB}[c][c]{\tiny{$\log$(dens)}}
\begin{minipage}[b]{0.21\linewidth}
\psfrag{AxA}[c][c]{\tiny{logit(\texttt{cfr})}}
\centering
\includegraphics[width=3cm, angle=270]{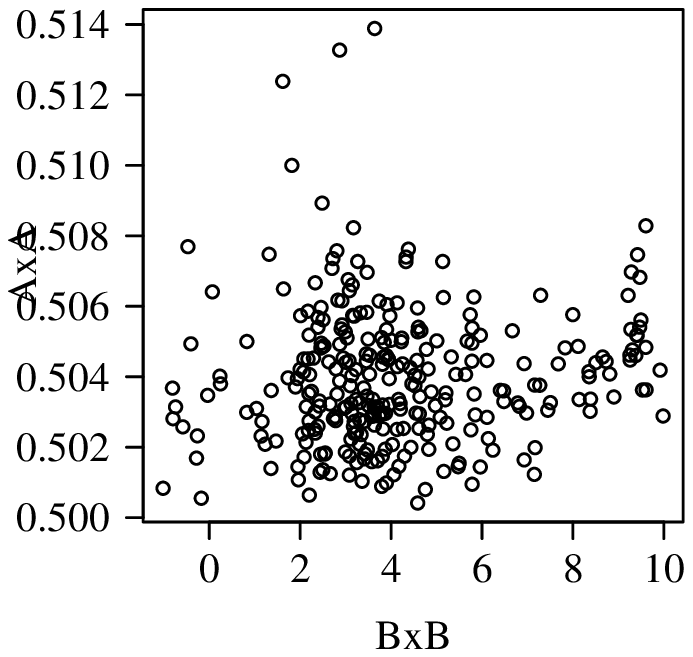}
\end{minipage} 
\hspace{0.2cm}
\begin{minipage}[b]{0.21\linewidth}
\psfrag{0.58}[c][c]{\tiny{}}
\psfrag{0.56}[c][c]{\tiny{}}
\psfrag{0.54}[c][c]{\tiny{}}
\psfrag{0.52}[c][c]{\tiny{0.52}}
\psfrag{AxA}[c][c]{\tiny{probit(\texttt{cfr})}}
\centering
\includegraphics[width=3cm, angle=270]{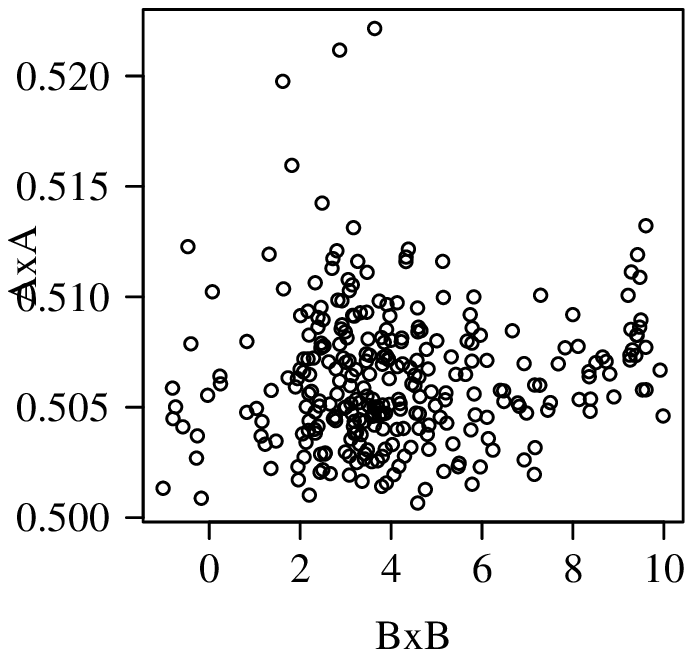}
\end{minipage}
\hspace{0.2cm}
\begin{minipage}[b]{0.21\linewidth}
\psfrag{AxA}[c][c]{\tiny{loglog(\texttt{cfr})}}
\centering
\includegraphics[width=3cm, angle=270]{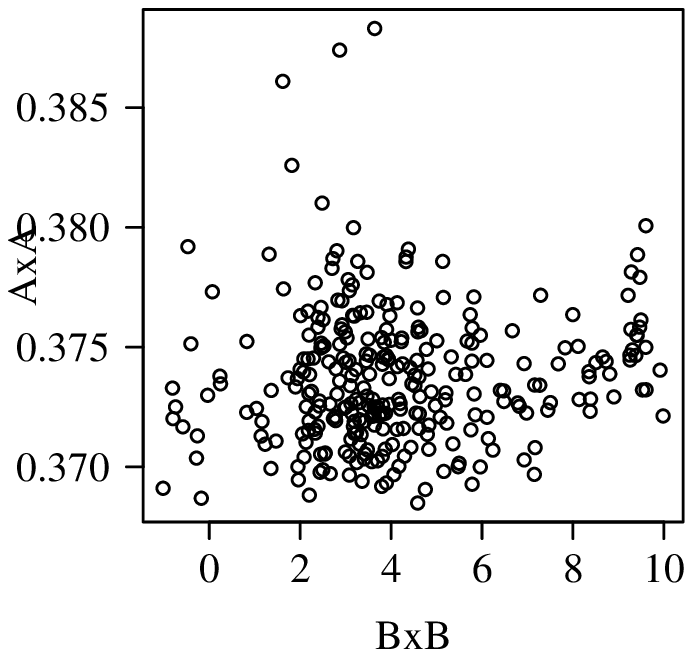}
\end{minipage} 
\hspace{0.2cm}
\begin{minipage}[b]{0.21\linewidth}
\psfrag{AxA}[c][c]{\tiny{cloglog(\texttt{cfr})}}
\centering
\includegraphics[width=3cm, angle=270]{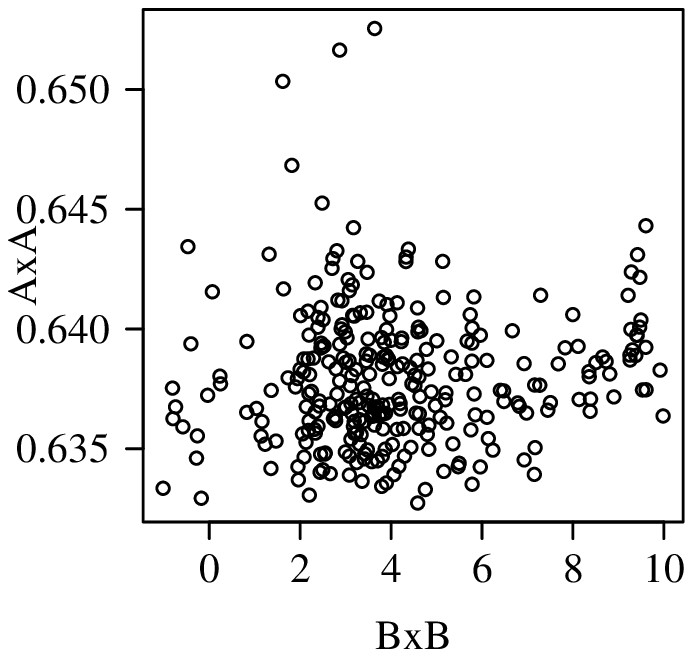}
\end{minipage} \\
\psfrag{BxB}[c][c]{\tiny{posit}}
\begin{minipage}[b]{0.21\linewidth}
\centering
\psfrag{AxA}[c][c]{\tiny{logit(\texttt{cfr})}}
\includegraphics[width=3cm, angle=270]{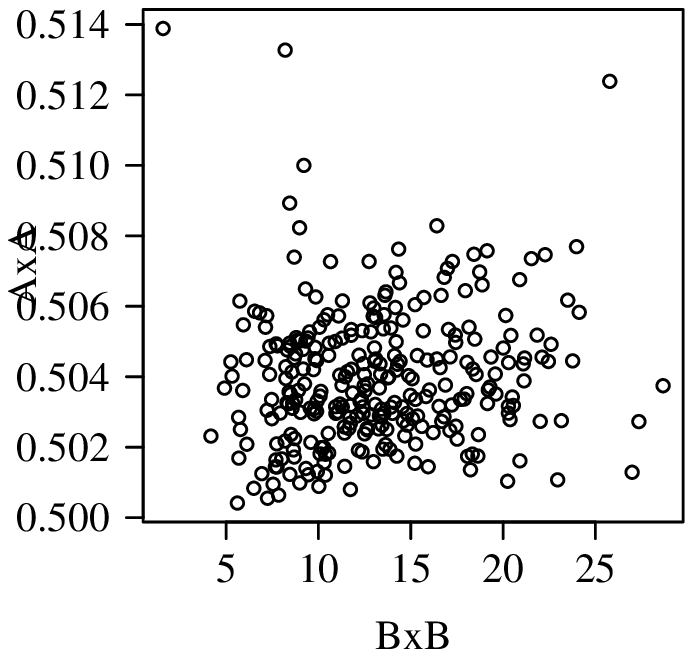}
\end{minipage} 
\hspace{0.2cm}
\begin{minipage}[b]{0.21\linewidth}
\centering
\psfrag{AxA}[c][c]{\tiny{probit(\texttt{cfr})}}
\includegraphics[width=3cm, angle=270]{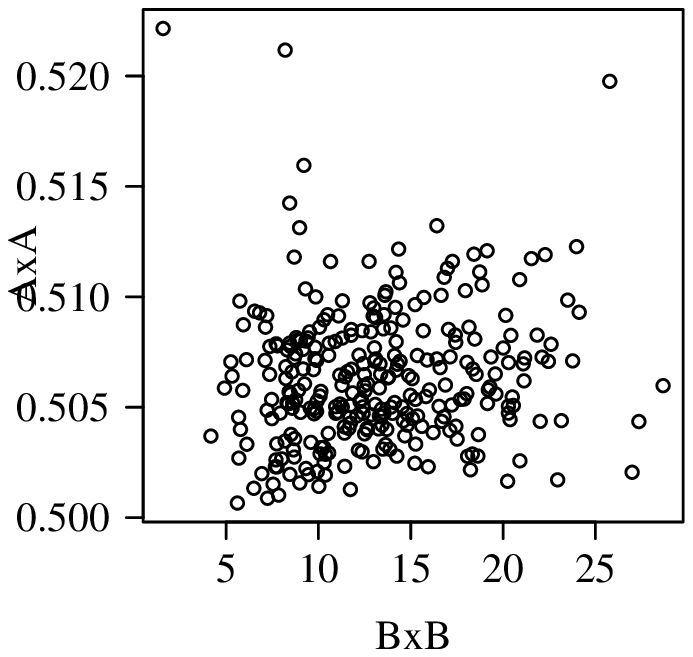}
\end{minipage}
\hspace{0.2cm}
\begin{minipage}[b]{0.21\linewidth}
\centering
\psfrag{AxA}[c][c]{\tiny{loglog(\texttt{cfr})}}
\includegraphics[width=3cm, angle=270]{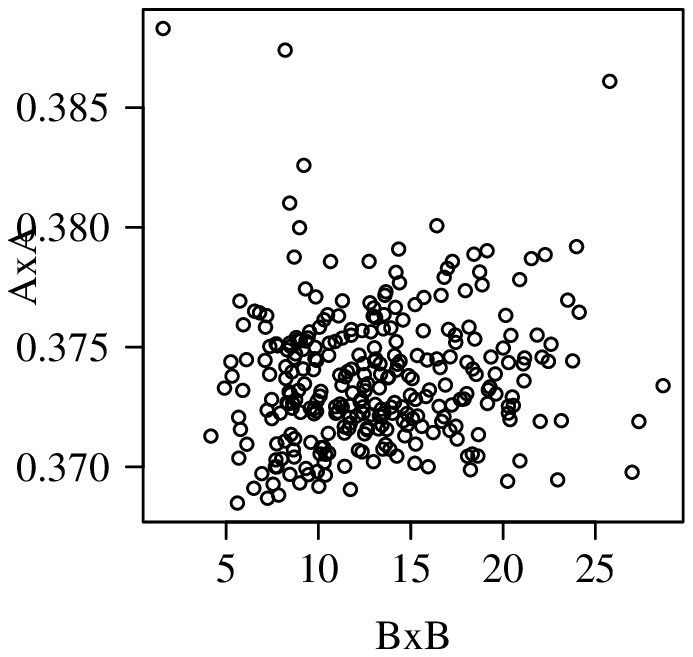}
\end{minipage} 
\hspace{0.2cm}
\begin{minipage}[b]{0.21\linewidth}
\centering
\psfrag{AxA}[c][c]{\tiny{cloglog(\texttt{cfr})}}
\includegraphics[width=3cm, angle=270]{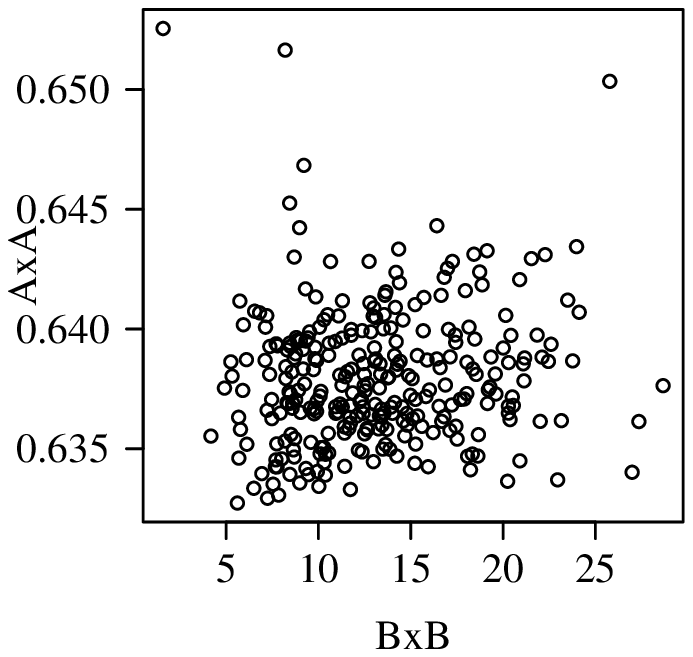}
\end{minipage} \\
\psfrag{BxB}[c][c]{\tiny{vaccine}}
\begin{minipage}[b]{0.21\linewidth}
\centering
\psfrag{AxA}[c][c]{\tiny{logit(\texttt{cfr})}}
\includegraphics[width=3cm, angle=270]{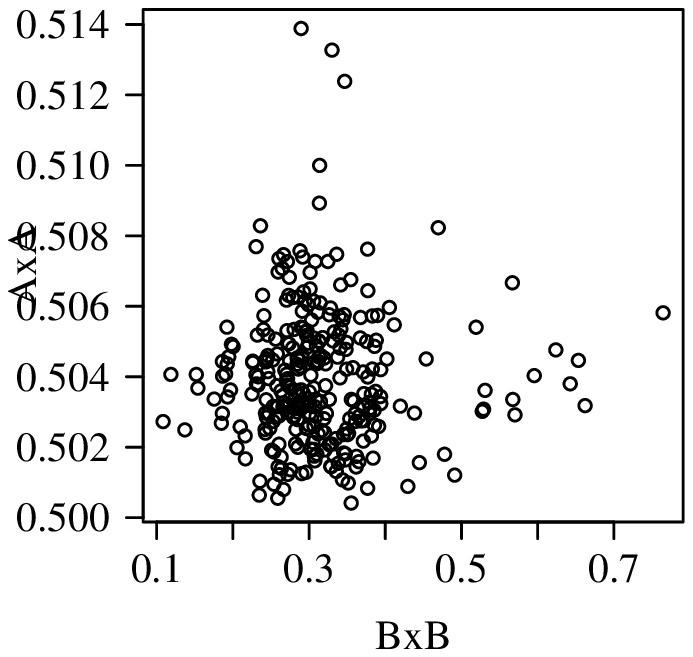}
\end{minipage} 
\hspace{0.2cm}
\begin{minipage}[b]{0.21\linewidth}
\centering
\psfrag{AxA}[c][c]{\tiny{probit(\texttt{cfr})}}
\includegraphics[width=3cm, angle=270]{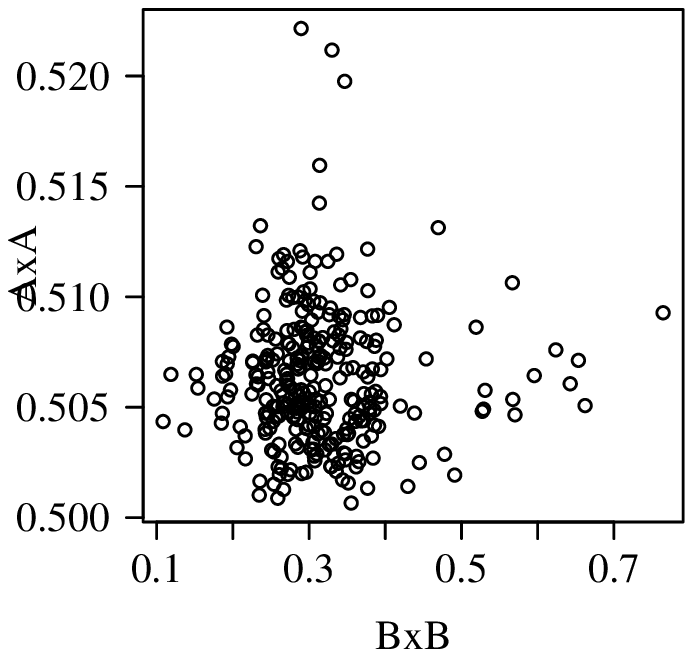}
\end{minipage}
\hspace{0.2cm}
\begin{minipage}[b]{0.21\linewidth}
\centering
\psfrag{AxA}[c][c]{\tiny{loglog(\texttt{cfr})}}
\includegraphics[width=3cm, angle=270]{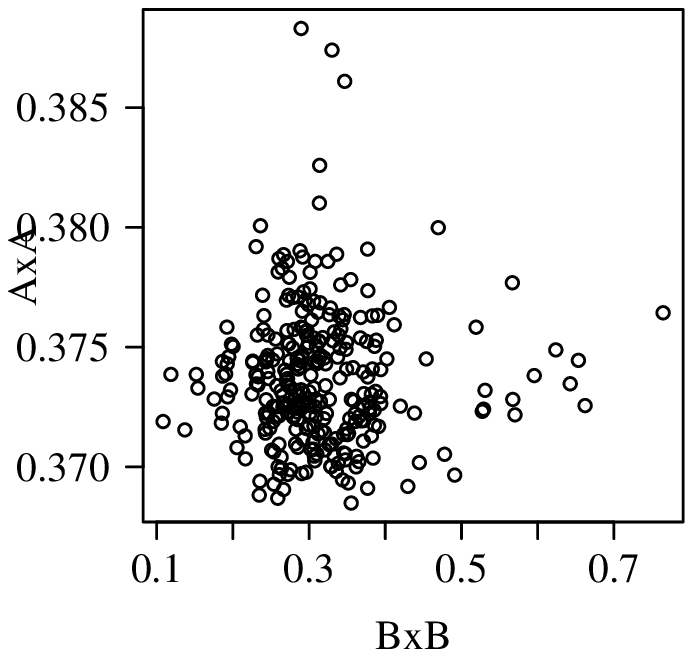}
\end{minipage} 
\hspace{0.2cm}
\begin{minipage}[b]{0.21\linewidth}
\centering
\psfrag{AxA}[c][c]{\tiny{cloglog(\texttt{cfr})}}
\includegraphics[width=3cm, angle=270]{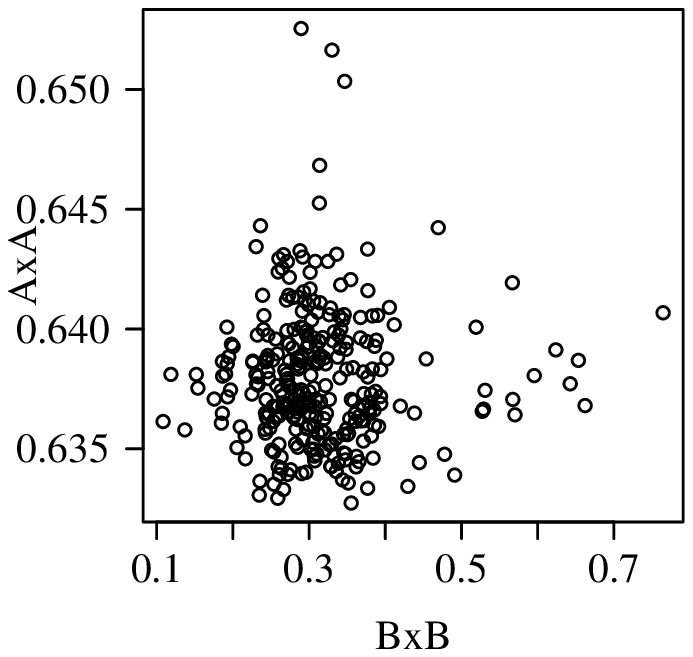}
\end{minipage} \\
\caption{Descriptive plots for $g_1$(\texttt{cfr}) versus log(\texttt{dens}), \texttt{posit}, and \texttt{vaccine} for different link functions: logit, probit, loglog, and cloglog.}\label{desc.COVID}
\end{figure}

Chile is administratively divided into 348 communes, which belong to 16 regions. We considered the cases and deaths of COVID-19 reported by the Chilean Ministry of Health for each commune between May 24 and July 23, 2021, but only for communes with at least 50 cases and 1 death, totaling 335,727 cases and 8,692 deaths in 305 communes. The communes were divided into three zones: North, Center, and South (see Figure \ref{figchile}). This division is typically used in Chile and is supported by climatological characteristics from the different zones. We also considered the following information for each commune:
\begin{itemize}
\item \texttt{cfr}: CFR (confirmed deaths/confirmed cases) between May 24 and July 23, 2021. Mean=0.026, Median=0.025, standard deviation=0.012, minimum=0.004, and maximum=0.099; see Figure \ref{figchile2}.
\item \texttt{dens}: population density of the commune (in km$^2$) according to the projected population of the Statistics National Institute of Chile for 2020; see Figure \ref{figchile3}.
\item \texttt{posit}: average of positive tests between January 25 and May 24, 2021; see Figure \ref{figchile3}.
\item \texttt{vaccine}: percentage of the population fully vaccinated by May 24, 2021; see Figure \ref{figchile3}.
\end{itemize}
Note that we used average \texttt{posit} during the last four months and \texttt{vaccine} up to May, 24 and the average \texttt{cfr} in the posterior two months from that date to model the quantiles of the average \texttt{cfr} for the next two months. Figure \ref{figchile2} suggests that the communes in the Center are more susceptible to have greater CFR than those in the North and South in the specified period. Additionally, based on Figure \ref{figchile3}, we also show that the average positivity rate in the four previous months was greater in communes of the Center, but the completed vaccination process, in general, was delayed in the North in comparison with the Center and South. Finally, the Center is the densest zone, especially in the Metropolitan (RM), V, and VIII regions.\\
We were interested in modeling the quantiles of the average CFR in terms of \texttt{dens}, \texttt{posit}, and \texttt{vaccine}. Figure \ref{desc.COVID} shows the plots for $g_1(\texttt{cfr})$ using different links versus $\log(\texttt{dens})$, $\texttt{posit}$, and $\texttt{vaccine}$. Note that, for this particular problem, the four links provided similar results.

\subsection{Inference}
\noindent

We considered modeling the $\tau$-th quantile of the CFR, say \texttt{cfr}$(\tau)$, such that $\texttt{cfr}_i(\tau)\sim \mbox{GB3}(\mu_i(\tau),\alpha_i(\tau),\beta_i(\tau))$, for $i=1, \ldots, n$, with
\begin{align*}
g_1(\mu_i(\tau))&=\theta_0(\tau)+\theta_1(\tau)\log(\texttt{dens}_i)+\theta_2(\tau)\texttt{posit}_i+\theta_3(\tau)\texttt{vaccine}_i, \nonumber \\
g_2(\alpha_i(\tau))&=\nu_0(\tau)+\nu_1(\tau)\log(\texttt{dens}_i)+\nu_2(\tau)\texttt{posit}_i+\nu_3(\tau)\texttt{vaccine}_i, \nonumber \\
g_3(\beta_i(\tau))&=\eta_0(\tau)+\eta_1(\tau)\log(\texttt{dens}_i)+\eta_2(\tau)\texttt{posit}_i+\eta_3(\tau)\texttt{vaccine}_i. \label{cov.in3}
\end{align*}

We considered the quantiles $\{0.1, 0.25, 0.50, 0.75, 0.90\}$ and used the four link functions previously discussed (logit, probit, loglog, and cloglog). For each combination of $\tau$ and link function, we selected the covariates based on the method discussed in Section \ref{cov.selection}. Considering that the terms related to the intercept, $\log$(\texttt{dens}), \texttt{posit}, and \texttt{vaccine} are variables 0 to 3, results of AIC and BIC criteria and selected covariates for each component are presented in Table \ref{AICBIC.app}. Note that the values of AIC and BIC are similar for the four considered links. Thus, we used the logit link (because it is the traditional link used) to continue the analysis. In addition, based on the results, the final model was
\begin{eqnarray*}
g_1(\mu_i(\tau))&=&\theta_0(\tau)+\theta_1(\tau)\log(\texttt{dens}_i)+\theta_3(\tau)\texttt{vaccine}_i, \nonumber \\
g_2(\alpha_i(\tau))&=&\nu_0(\tau)+\nu_1(\tau)\log(\texttt{dens}_i)+\nu_2(\tau)\texttt{posit}_i, \nonumber \\
g_3(\beta_i(\tau))&=&\eta_0(\tau) \nonumber
\label{cov.in3}.
\end{eqnarray*}
In other words, the quantile was modeled by $\log(\texttt{dens})$ and $\texttt{vaccine}$, $\alpha$ was modeled by $\log(\texttt{dens})$ and \texttt{posit}, whereas $\beta$ only considered the intercept.

\begin{table}
\centering
\caption{AIC, BIC, and selected covariates for the COVID-19 dataset under different link functions for the modeled quantile.}\label{AICBIC.app}
\begin{tabular}{llllccc}
\hline
$\tau$      & link    & AIC & BIC & $\mu$ & $\alpha$ & $\beta$ \\
\hline
\multirow{4}{*}{$0.1$} & logit &  $-$1945.6  &  $-$1919.6   & 013   &  012    &  0    \\
                     & probit  &  $-$1945.7  &  $-$1919.7   & 013   &  012    &  0    \\
                     & loglog  &  $-$1945.8  &  $-$1919.7   & 013   &  012    &  0    \\
                     & cloglog &  $-$1945.6  &  $-$1919.6   & 013   &  012    &  0    \\ \hline
\multirow{4}{*}{$0.25$} & logit & $-$1947.5  &  $-$1921.5   & 013   &  012    &  0    \\
                     & probit   & $-$1947.4  &  $-$1921.4   & 013   &  012    &  0    \\
                     & loglog   & $-$1947.3  &  $-$1921.2   & 013   &  012    &  0    \\
                     & cloglog  & $-$1947.5  &  $-$1921.5   & 013   &  012    &  0    \\ \hline
\multirow{4}{*}{$0.5$} & logit &  $-$1945.4  &  $-$1919.4   & 013   &  012    &  0    \\
                     & probit  &  $-$1945.3  &  $-$1919.3   & 013   &  012    &  0    \\
                     & loglog  &  $-$1945.2  &  $-$1919.1   & 013   &  012    &  0    \\
                     & cloglog &  $-$1945.5  &  $-$1919.4   & 013   &  012    &  0    \\ \hline
\multirow{4}{*}{$0.75$} & logit & $-$1940.2  &  $-$1914.1   & 013   &  012    &  0    \\
                     & probit  &  $-$1940.1  &  $-$1914.1   & 013   &  012    &  0    \\
                     & loglog  &  $-$1940.1  &  $-$1914.0   & 013   &  012    &  0    \\
                     & cloglog &  $-$1940.2  &  $-$1914.1   & 013   &  012    &  0    \\ \hline
\multirow{4}{*}{$0.9$} & logit &  $-$1936.8  &  $-$1918.2   & 03    &  01     &  0    \\
                     & probit  &  $-$1936.8  &  $-$1918.2   & 03    &  01     &  0    \\
                     & loglog  &  $-$1936.8  &  $-$1918.2   & 03    &  01     &  0    \\
                     & cloglog &  $-$1936.8  &  $-$1918.2   & 03    &  01     &  0    \\ \hline
\end{tabular}
\end{table}

\begin{figure}[!htbp]
\psfrag{AxA}[c][c]{\scriptsize{$\tau$}}
\psfrag{0.2}[c][c]{\tiny{0.2}}
\psfrag{0.4}[c][c]{\tiny{0.4}}
\psfrag{0.6}[c][c]{\tiny{0.6}}
\psfrag{0.8}[c][c]{\tiny{0.8}}
\begin{minipage}[b]{0.21\linewidth}
\psfrag{BxB}[c][c]{\scriptsize{$\theta_{0}(\tau)$}}
\psfrag{-5.0}[c][c]{\tiny{-5.0}}
\psfrag{-4.5}[c][c]{\tiny{-4.5}}
\psfrag{-4.0}[c][c]{\tiny{-4.0}}
\psfrag{-3.5}[c][c]{\tiny{-3.5}}
\centering
\includegraphics[width=3cm, angle=270]{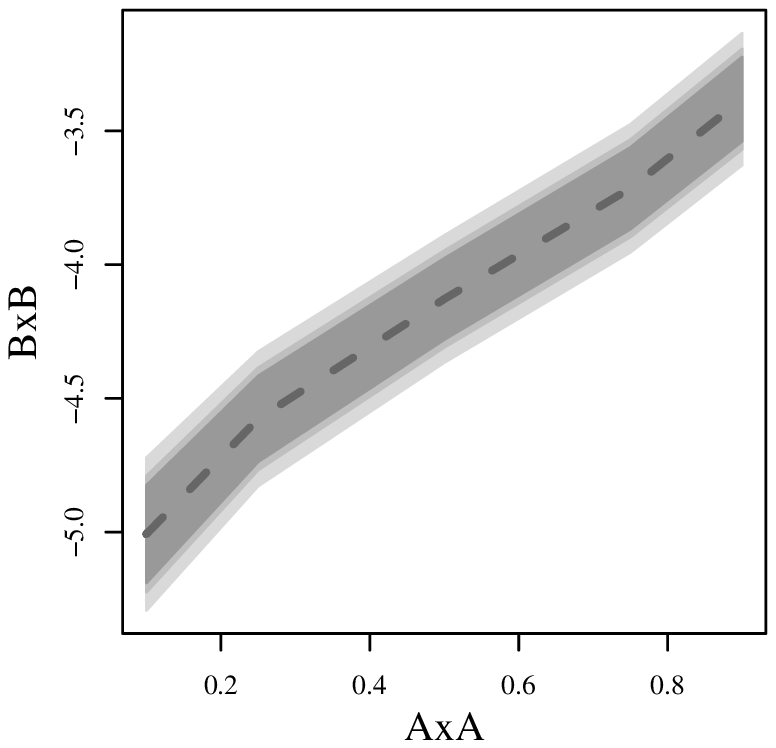}
\end{minipage} 
\hspace{0.2cm}
\begin{minipage}[b]{0.21\linewidth}
\psfrag{BxB}[c][c]{\scriptsize{$\theta_{1}(\tau)$}}
\psfrag{0.00}[c][c]{\tiny{0.00}}
\psfrag{0.05}[c][c]{\tiny{0.05}}
\psfrag{0.10}[c][c]{\tiny{0.10}}
\centering
\includegraphics[width=3cm, angle=270]{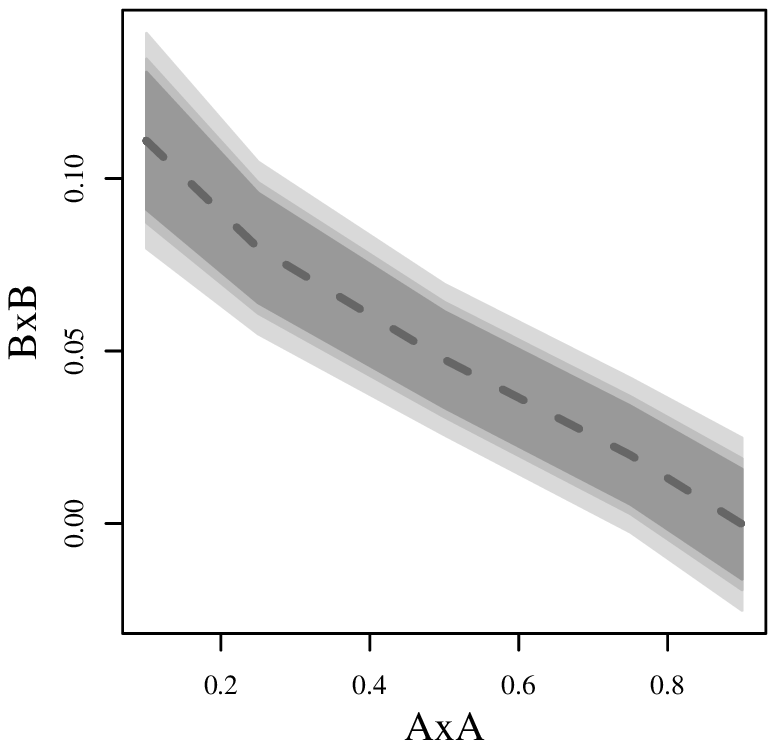}
\end{minipage}
\hspace{0.2cm}
\begin{minipage}[b]{0.21\linewidth}
\psfrag{BxB}[c][c]{\scriptsize{$\theta_{3}(\tau)$}}
\psfrag{0.000}[c][c]{\tiny{0.00}}
\psfrag{0.010}[c][c]{\tiny{0.01}}
\psfrag{0.020}[c][c]{\tiny{0.02}}
\psfrag{0.030}[c][c]{\tiny{0.03}}
\centering
\includegraphics[width=3cm, angle=270]{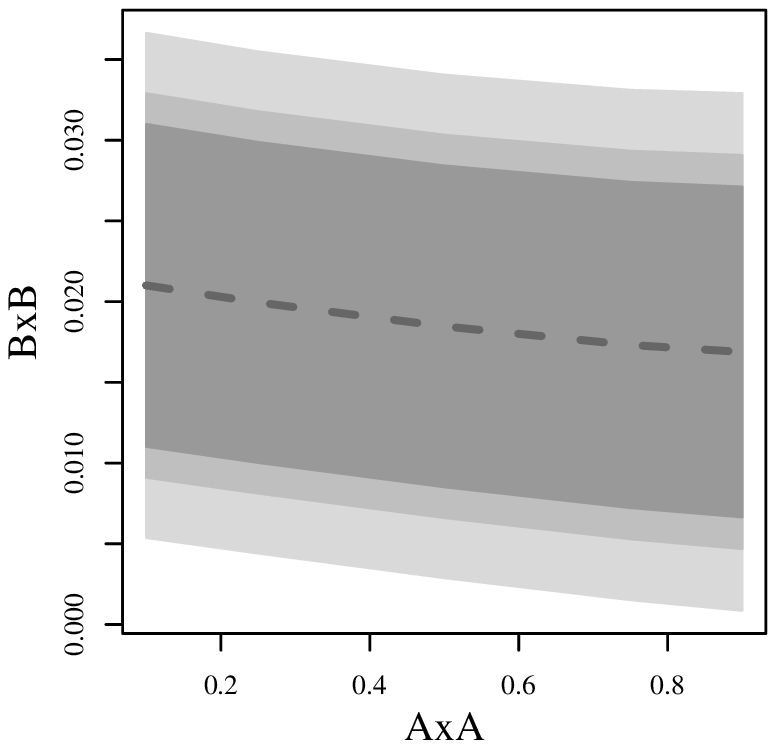}
\end{minipage} 
\hspace{0.2cm}
\begin{minipage}[b]{0.21\linewidth}
\psfrag{BxB}[c][c]{\scriptsize{$\nu_{0}(\tau)$}}
\psfrag{0.5}[c][c]{\tiny{0.5}}
\psfrag{1.0}[c][c]{\tiny{1.0}}
\psfrag{1.5}[c][c]{\tiny{1.5}}
\psfrag{2.0}[c][c]{\tiny{2.0}}
\psfrag{2.5}[c][c]{\tiny{2.5}}
\centering
\includegraphics[width=3cm, angle=270]{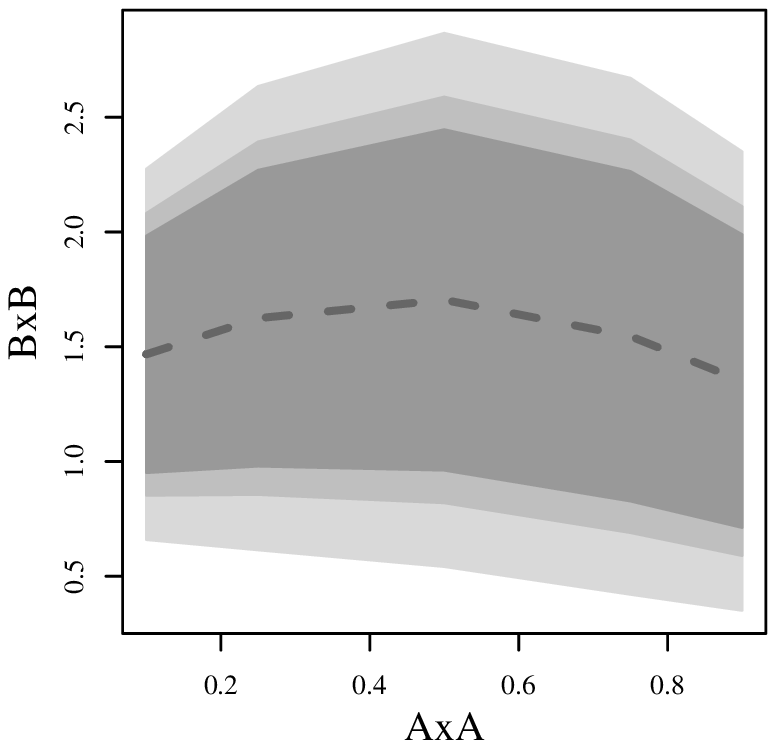}
\end{minipage} 
\\
\begin{minipage}[b]{0.21\linewidth}
\psfrag{BxB}[c][c]{\scriptsize{$\nu_{1}(\tau)$}}
\psfrag{0.2}[c][c]{\tiny{0.2}}
\psfrag{0.3}[c][c]{\tiny{0.3}}
\psfrag{0.4}[c][c]{\tiny{0.4}}
\psfrag{0.5}[c][c]{\tiny{0.5}}
\centering
\includegraphics[width=3cm, angle=270]{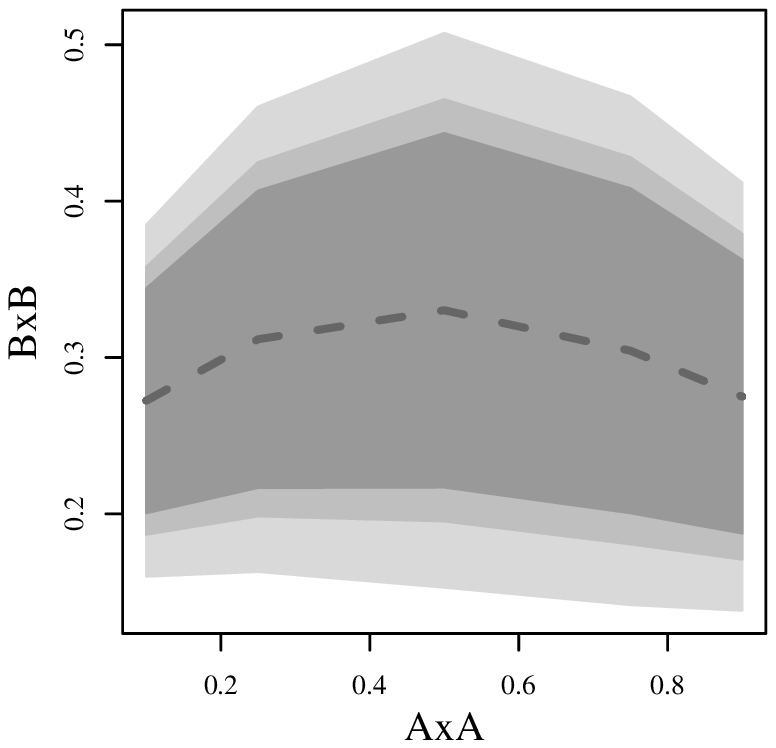}
\end{minipage}
\hspace{0.2cm}
\begin{minipage}[b]{0.21\linewidth}
\psfrag{BxB}[c][c]{\scriptsize{$\nu_{2}(\tau)$}}
\psfrag{-5}[c][c]{\tiny{-5}}
\psfrag{-4}[c][c]{\tiny{-4}}
\psfrag{-3}[c][c]{\tiny{-3}}
\psfrag{-2}[c][c]{\tiny{-2}}
\psfrag{-1}[c][c]{\tiny{-1}}
\psfrag{0}[c][c]{\tiny{0}}
\centering
\includegraphics[width=3cm, angle=270]{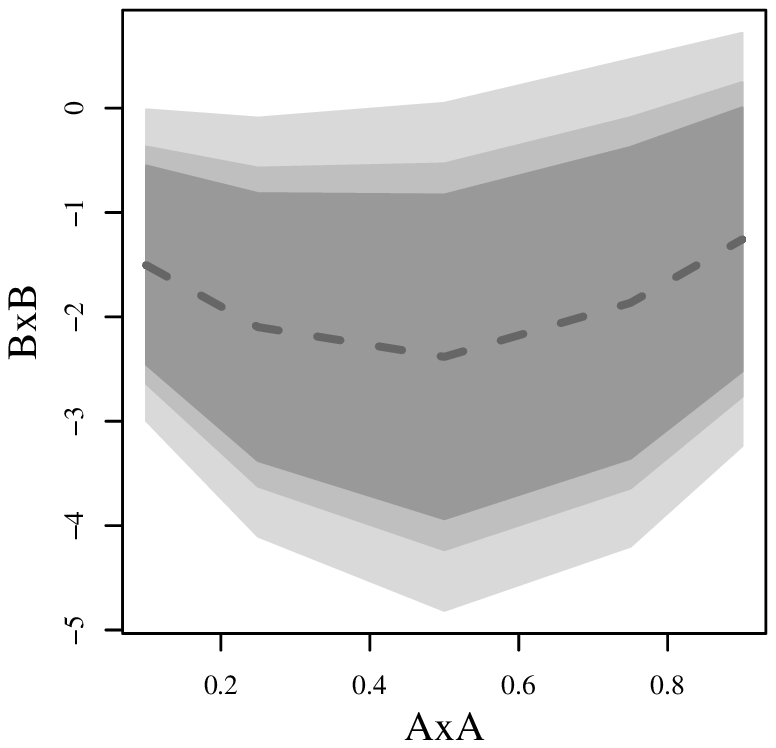}
\end{minipage} 
\hspace{0.2cm}
\begin{minipage}[b]{0.21\linewidth}
\psfrag{BxB}[c][c]{\scriptsize{$\eta_{0}(\tau)$}}
\psfrag{2}[c][c]{\tiny{2}}
\psfrag{3}[c][c]{\tiny{3}}
\psfrag{4}[c][c]{\tiny{4}}
\psfrag{5}[c][c]{\tiny{5}}
\psfrag{6}[c][c]{\tiny{6}}
\centering
\includegraphics[width=3cm, angle=270]{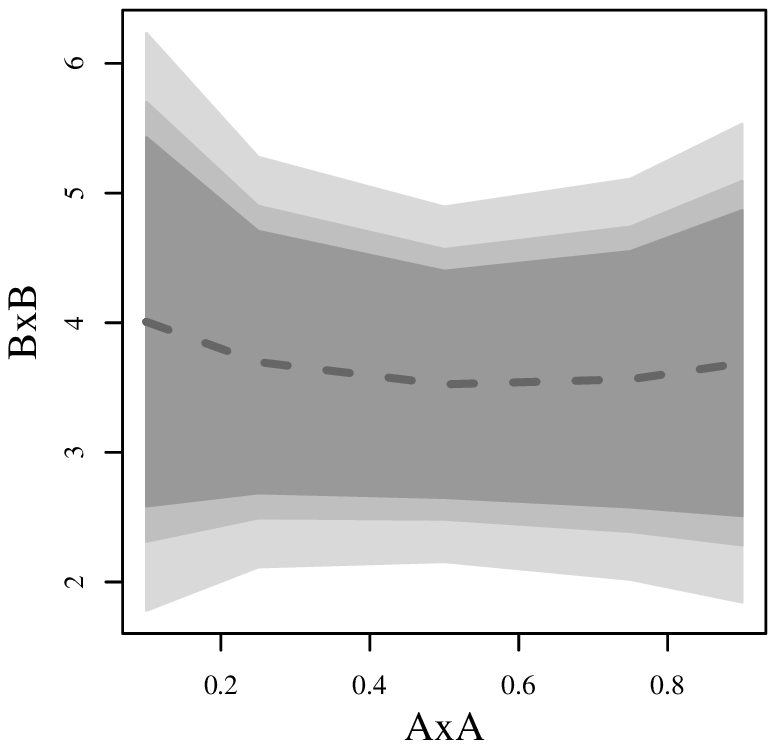}
\end{minipage} 
\caption{COVID-19 data set: Point estimates and 90\%, 95\%, and 99\% confidence interval (CI) for model parameters under the GB3 quantile regression model.}
\label{profilelike}
\end{figure}

\begin{figure}[!htbp]
\centering
\psfrag{-3}[c][c]{\scriptsize{-3}}
\psfrag{-2}[c][c]{\scriptsize{-2}}
\psfrag{-1}[c][c]{\scriptsize{-1}}
\psfrag{0}[c][c]{\scriptsize{0}}
\psfrag{1}[c][c]{\scriptsize{1}}
\psfrag{2}[c][c]{\scriptsize{2}}
\psfrag{3}[c][c]{\scriptsize{3}}
\psfrag{4}[c][c]{\scriptsize{4}}
\psfrag{BxB}[c][c]{\scriptsize{\texttt{sample quantile}}}
\psfrag{AxA}[c][c]{\scriptsize{\texttt{N(0,1) quantile}}}
\begin{minipage}[b]{0.30\linewidth}
\centering
\includegraphics[width=4cm, angle=270]{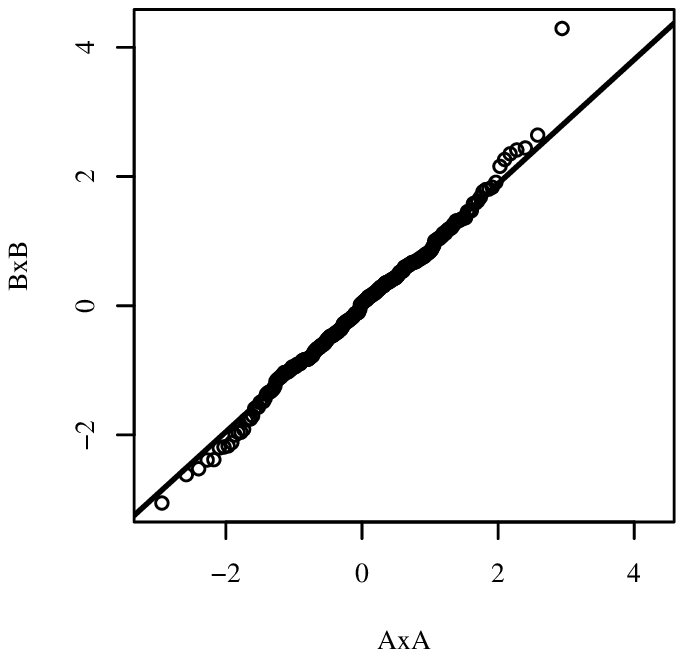}
\end{minipage} 
\hspace{0.1cm}
\begin{minipage}[b]{0.30\linewidth}
\centering
\includegraphics[width=4cm, angle=270]{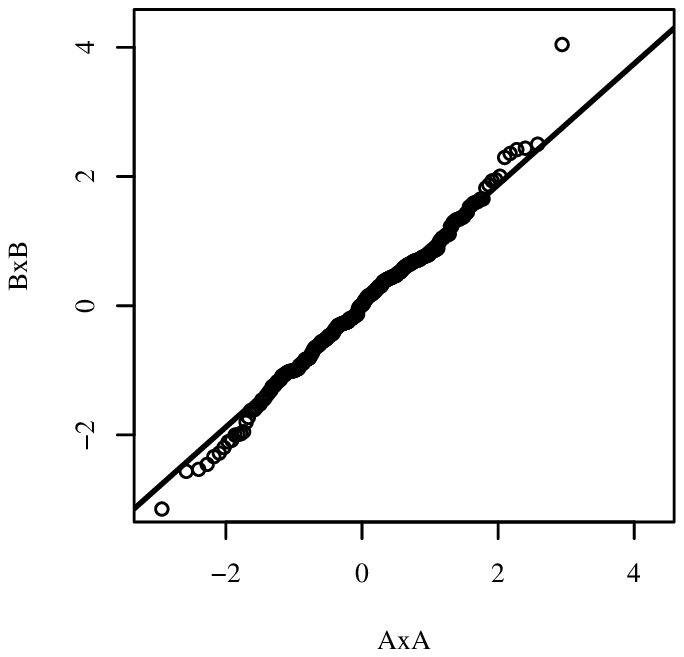}
\end{minipage}
\hspace{0.1cm}
\begin{minipage}[b]{0.30\linewidth}
\centering
\includegraphics[width=4cm, angle=270]{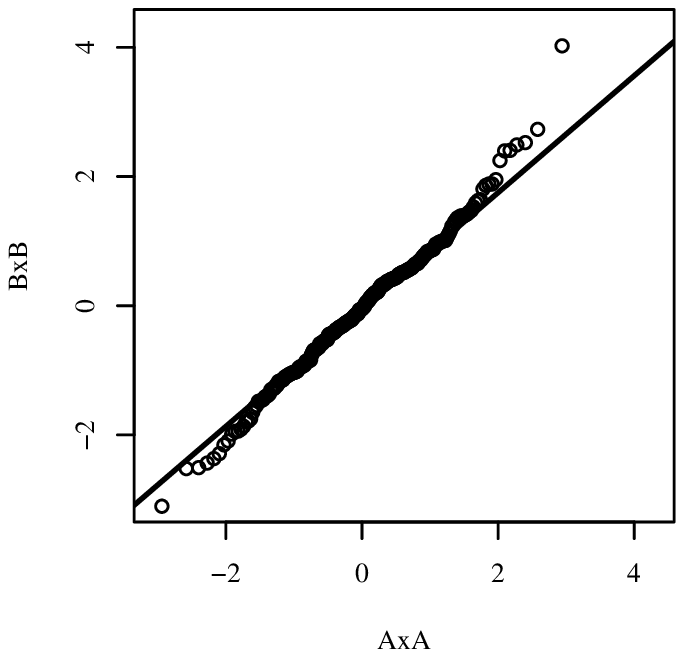}
\end{minipage} 
\caption{QQ plot of RQ residuals for the GB3 regression model for different quantiles:  $\tau=0.10$ (left panel), $\tau=0.50$ (center panel), and $\tau=0.90$ (right panel); COVID-19 dataset in Chile.}
\label{profilelike2}
\end{figure}

\begin{figure}[!htbp]
\begin{center}
\resizebox{\linewidth}{!}{
\begin{tabular}{cccc}
North & Center & South \\
  \begin{minipage}{.28\textwidth}{\includegraphics[width=1.3\textwidth]{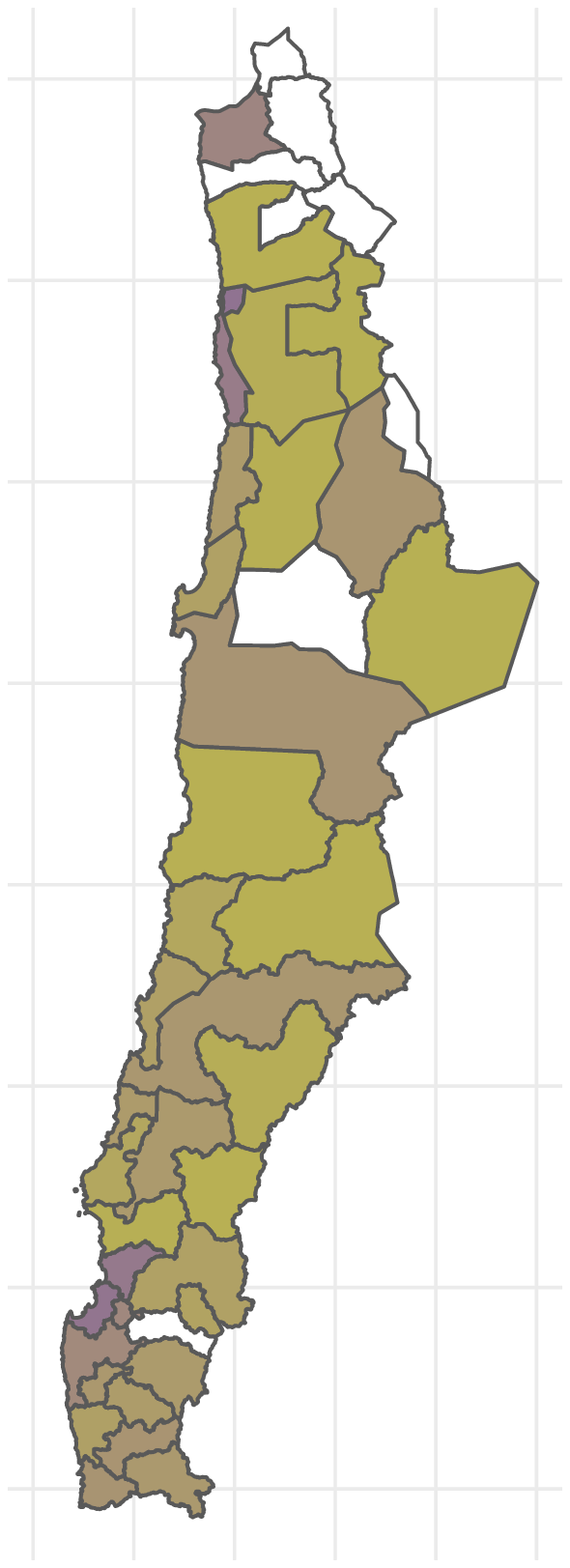}} \end{minipage} &
  \begin{minipage}{.28\textwidth}{\includegraphics[width=1.3\textwidth]{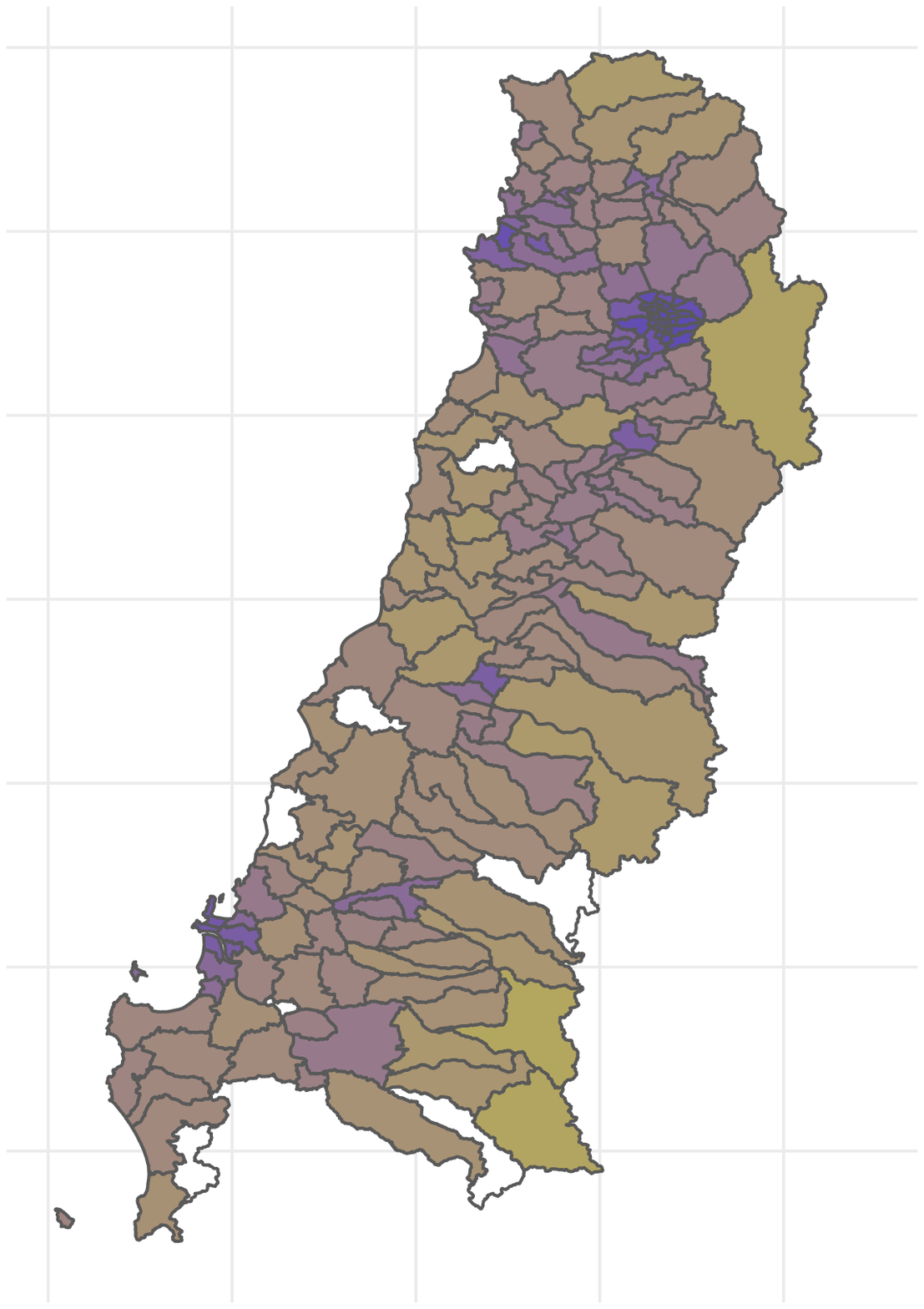}} \end{minipage} &
  \begin{minipage}{.28\textwidth}{\includegraphics[width=1.3\textwidth]{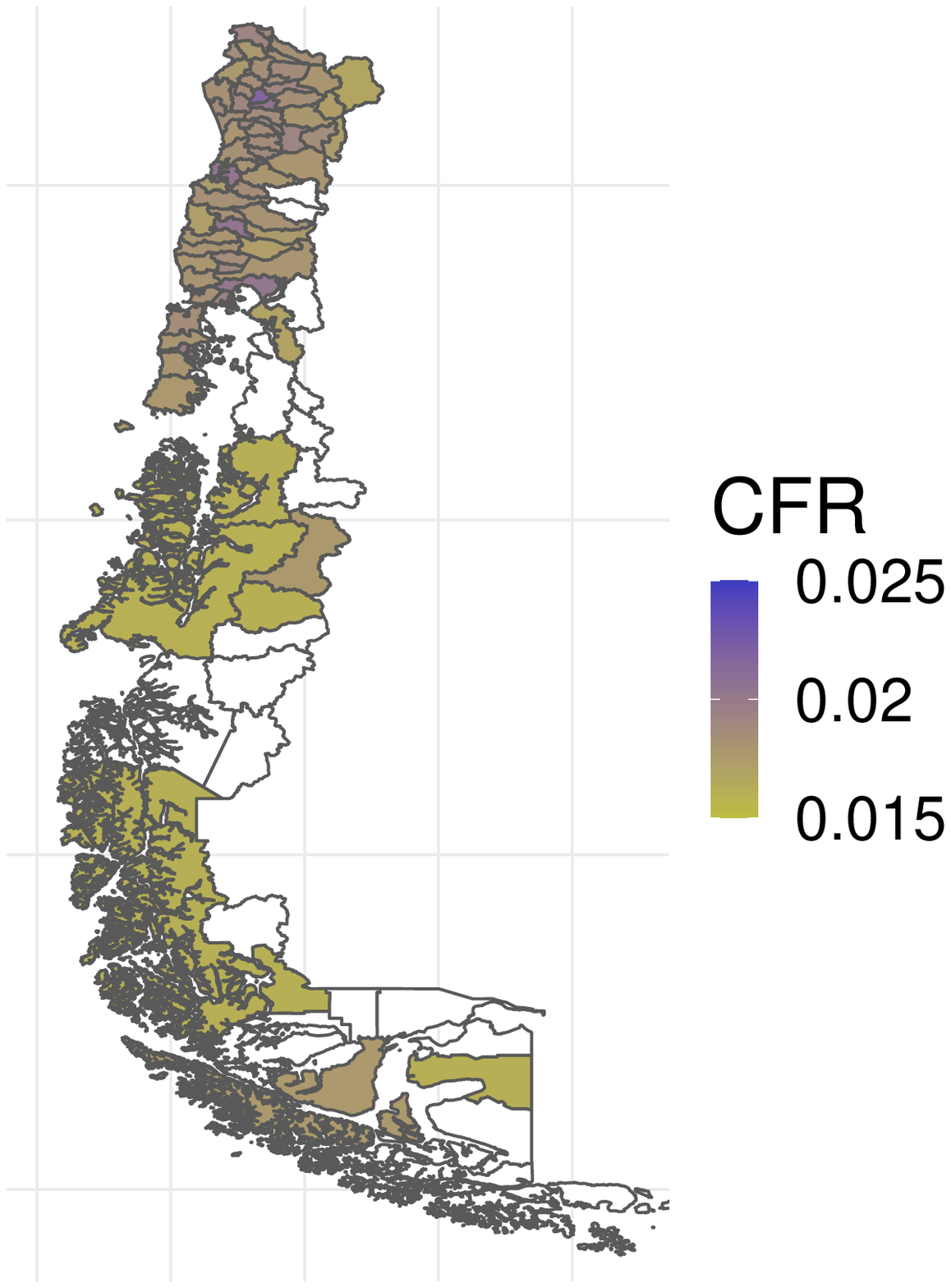}} \end{minipage} \\
\end{tabular}
}
\end{center}
\caption{Estimated median for the average CFR in the two next months in the communes of the North, Center, and South zones of Chile if the vaccination process was augmented in 30\%.}
\label{figchile.est}
\end{figure}

Table \ref{tab:resulall} presents the ML estimates and standard errors (SEs) for the GB3 quantile regression model parameters with $\tau = \{0.10, 0.25, 0.50, 0.75, 0.90\}$. Figure \ref{profilelike} plots the estimated parameters of the proposed model across such quantiles. We observe that the estimates of $\theta_{0}(\tau)$ increase across $\tau$, implying that the $\tau$-th quantile of CFR increases as $\tau$ increases for the case when $\log$(\texttt{dens}) and \texttt{vaccine} are fixed at 0. On the other hand, the estimates of $\theta_{1}(\tau)$ decrease across $\tau$, i.e., the $\tau$-th quantile of CFR decreases as \texttt{ldens} increases. This is expected because denser cities have greater and better access to health centers. Similarly, we can conclude that the $\tau$-th quantile of CFR decreases as \texttt{vaccine} increases. Note that from the discussion provided in Equation (\ref{interp}), as $1-\exp(-\theta_1(\tau=0.5)/2)=-0.023$ and $1-\exp(\theta_3(\tau=0.5)/2)=-0.009$, we also concluded that the median of the average CFR:
\begin{itemize}
\item decreases by approximately 2.3\% when $\log$(\texttt{dens}) is increased by one unit and vaccination is fixed;
\item decreases by approximately 0.9\% when the vaccination is increased by 1\% in the population and $\log$(\texttt{dens}) is fixed.
\end{itemize}
Additionally, considering the estimates for $\tau=0.5$, we have that $\exp(\nu_1)=\exp(0.3302)-1=0.391$.
Therefore, the deaths related to COVID-19 in Chile in the two last months increase by $39.1\%$ if the $\log$(\texttt{dens}) is increased by one unit. 
On the other hand, as the $\beta$ parameter is constant, variations in density, vaccination, and positivity do not produce significant variations in the total COVID-19 cases in Chile.
Figure~\ref{profilelike2} shows the QQ plots of the RQ residuals for the GB3 quantile regression model in Table \ref{tab:resulall}. The proposed model provided a satisfactory fit to the data as most empirical quantiles agree with the theoretical ones.
Finally, Figure \ref{figchile.est} shows the predicted median in the two last months for the CFR in the three zones if the vaccination process is increased by 30\% in each commune (capped at 100\%).
Comparing Figures \ref{figchile2} and \ref{figchile.est}, we observe a large decrease in the predicted median for the average CFR in the two last months, showing that the vaccination is an effective way to reduce COVID-19 deaths.

\begin{table}[!htbp]
\centering
\caption{Estimated parameters and SEs (given in parentheses) for the GB3 quantile regression model using the logit link for the COVID-19 dataset.}
\label{tab:resulall}
\begin{center}
\resizebox{\linewidth}{!}{
\begin{tabular}{crrrrr}
\hline
           & \multicolumn{5}{c}{Quantile}            \\
 Parameter          &   $\tau=0.10$ &       $\tau=0.25$ &   $\tau=0.50$ &       $\tau=0.75$ &   $\tau=0.90$  \\
\hline

$\theta_{0}(\tau)$ & $-$5.0074 (0.1108) & $-$4.5767 (0.0975) & $-$4.1285 (0.0923) & $-$3.7146 (0.0928) & $-$3.3814 (0.0957) \\

$\theta_{1}(\tau)$ & 0.1109 (0.0121) & 0.0799 (0.0097) & 0.0475 (0.0085) & 0.0198 (0.0087) & $-$0.0002 (0.0097) \\

$\theta_{3}(\tau)$ & 0.0210 (0.0061) & 0.0200 (0.0060) & 0.0185 (0.0060) & 0.0173 (0.0061) & 0.0169 (0.0062) \\
\hline

$\nu_{0}(\tau)$ & 1.4671 (0.3136) & 1.6245 (0.3921) & 1.7041 (0.4512) & 1.5457 (0.4366) & 1.3505 (0.3880) \\

$\nu_{1}(\tau)$ & 0.2724 (0.0436) & 0.3117 (0.0577) & 0.3302 (0.0688) & 0.3043 (0.0631) & 0.2750 (0.0531) \\

$\nu_{2}(\tau)$ & $-$1.5053 (0.5783) & $-$2.0980 (0.7778) & $-$2.3832 (0.9432) & $-$1.8658 (0.9055) & $-$1.2587 (0.7672) \\
\hline

$\eta_{0}(\tau)$ & 4.0068 (0.8642) & 3.6964 (0.6137) & 3.5249 (0.5311) & 3.5639 (0.5994) & 3.6883 (0.7157) \\

\hline

\end{tabular}
}
\end{center}
\end{table}

\newpage

\section{Discussion}\label{sec:6}
\noindent

In this paper, we proposed a new parametric quantile regression model for dealing with data in the interval $(0, 1)$. The proposed model is based on a reparameterization of the GB3 distribution, which has the distribution quantile as one of its parameters.
In particular, we proposed a regression model to describe data of the type $V/(W + V)$ at different quantiles.
The estimates of the model parameters have a simple interpretation in terms of percentage increases/decreases of the quantile $q(V/(W + V)|\mathbf{X})$. We allowed a regression structure for the $\mu$, $\alpha$, and $\beta$ parameters and used maximum likelihood inference to estimate
the model parameters, which can be easily computed using the \texttt{R} software. Therefore, our model required less computational cost. We carried out a Monte Carlo simulation study to evaluate the recovery of the parameters and assess the choice of the link functions. The numerical results showed a good performance of the maximum likelihood estimates. We also applied the proposed model to a real COVID-19 dataset from Chile. In particular, we modeled the case fatality rate (CFR) of COVID-19 and used the population density, positivity rate, and percentage of the population fully vaccinated as covariates. The main conclusions were that the median of the CFR in the two last months decreased by approximately 2.3\% when the logarithm of the population density was increased by one unit (a denser region has greater and better access to health centers) and decreased by approximately 0.9\% when the vaccination process was increased by 1\%. The results were quite favorable to the proposed GB3 quantile regression model and emphasized the importance of using a quantile approach. As part of future research, we envisage to propose zero-or-one inflated quantile regression models based on the GB3 model.

\normalsize


\begin{thebibliography}{}

\bibitem[Ahuja, 1969]{ahuja:69}
Ahuja, J.~C. (1969).
\newblock {On certain properties of the generalized Gompertz distribution}.
\newblock {\em Sankhyā}, B31:541--544.

\bibitem[Akaike, 1974]{akaike1974}
Akaike, H. (1974).
\newblock A new look at the statistical model identification.
\newblock {\em IEEE Transactions on Automatic Control}, 19:716--723.

\bibitem[Barr\'ia-Sandoval et~al., 2021]{BarriaSandoval:21}
Barr\'ia-Sandoval, C., Ferreira, G., Benz-Parra, K., and L\'opez-Flores, P.
  (2021).
\newblock {Prediction of confirmed cases of and deaths caused by COVID-19 in
  Chile through time series techniques: A comparative study}.
\newblock {\em PLOS ONE}, 16:e0245414.

\bibitem[Bayes et~al., 2017]{bayesetal:17}
Bayes, C.~L., Baz\'an, J.~L., and Castro, M. (2017).
\newblock {A quantile parametric mixed regression model forbounded response
  variables}.
\newblock {\em Statistics and Its Interface}, 10:483--493.

\bibitem[Berger et~al., 2019]{bergeretal:19}
Berger, M., Wagner, M., and Schmid, M. (2019).
\newblock Modeling biomarker ratios with gamma distributed components.
\newblock {\em The Annals of Applied Statistics}, 13.

\bibitem[Bottai et~al., 2010]{bottai10}
Bottai, M., Cai, B., and McKeown, R.~E. (2010).
\newblock Logistic quantile regression for bounded outcomes.
\newblock {\em Statistics in Medicine}, 29(2):309--317.

\bibitem[Cordeiro et~al., 2014]{cordeirosantanaetal:14}
Cordeiro, G.~M., de~Santana, L.~H., Ortega, E. M.~M., and Pescim, R.~R. (2014).
\newblock A new family of distributions: Libby-novick beta.
\newblock {\em International Journal of Statistics and Probability}, 3:63--80.

\bibitem[Cox and Hinkley, 1979]{Cox1979}
Cox, D.~R. and Hinkley, D.~V. (1979).
\newblock {\em Theoretical Statistics}.
\newblock CRC Press.

\bibitem[Dunn and Smyth, 1996]{ds:96}
Dunn, P. and Smyth, G. (1996).
\newblock {Randomized quantile residuals}.
\newblock {\em Journal of Computational and Graphical Statistics}, 5:236--244.

\bibitem[Ferrari and Cribari-Neto, 2004]{fcn:04}
Ferrari, S. and Cribari-Neto, F. (2004).
\newblock {Beta regression for modelling rates and proportions}.
\newblock {\em Journal of Applied Statistics}, 31:799--815.

\bibitem[Galarza et~al., 2017]{galarzaetal:17}
Galarza, C., Lachos, V.~H., Cabral, C. R.~B., and Castro, C.~L. (2017).
\newblock Robust quantile regression using a generalized class of skewed
  distributions.
\newblock {\em Stat}, 6:113--130.

\bibitem[Guerrero-Nancuante and Manr\'iquez, 2020]{Guerrero-Nancuante:20}
Guerrero-Nancuante, C. and Manr\'iquez, P. (2020).
\newblock {An epidemiological forecast of COVID-19 in Chile based on the
  generalized SEIR model and the concept of recovered}.
\newblock {\em Medwave}.

\bibitem[Hao and Naiman, 2007]{hn:07}
Hao, L. and Naiman, D. (2007).
\newblock {\em Quantile Regression}.
\newblock Sage Publications, California , US.

\bibitem[Koenker, 2005]{k:05}
Koenker, R. (2005).
\newblock {\em Quantile Regression}.
\newblock Cambridge University Press, Cambridge, UK.

\bibitem[Libby and Novick, 1982]{libbynovick:82}
Libby, D.~L. and Novick, M.~R. (1982).
\newblock Multivariate generalized beta-distributions with applications to
  utility assessment.
\newblock {\em Journal of Educational Statistics}, 7.

\bibitem[Malik, 1967]{malik:67}
Malik, H.~J. (1967).
\newblock {Exact distribution of the quotient of independent generalized gamma
  variables}.
\newblock {\em Canadian Mathematical Bulletin}, 10:463--465.

\bibitem[McDonald and Xu, 1995]{MC95}
McDonald, J.~B. and Xu, Y.~J. (1995).
\newblock A generalization of the beta distribution with applications.
\newblock {\em Journal of Econometrics}, 66(1):133--152.

\bibitem[Mittelhammer et~al., 2000]{mjm:00}
Mittelhammer, R.~C., Judge, G.~G., and Miller, D.~J. (2000).
\newblock {\em {Econometric Foundations}}.
\newblock Cambridge University Press, New York, US.

\bibitem[Noufaily and Jones, 2013]{nj:13}
Noufaily, A. and Jones, M. (2013).
\newblock {Parametric quantile regression based on the generalized gamma
  distribution}.
\newblock {\em Journal of the Royal Statistical Society C}, 62:723--740.

\bibitem[{R Core Team}, 2021]{R:21}
{R Core Team} (2021).
\newblock {\em {R: A Language and Environment for Statistical Computing}}.
\newblock R Foundation for Statistical Computing, Vienna, Austria.

\bibitem[Ristić et~al., 2015]{risticetal:15}
Ristić, M.~M., Popović, B.~V., and Nadarajah, S. (2015).
\newblock Libby and novick's generalized beta exponential distribution.
\newblock {\em Journal of Statistical Computation and Simulation}, 85:740--761.

\bibitem[S\'anchez et~al., 2021]{slgs:2020a}
S\'anchez, L., Leiva, V., Galea, M., and Saulo, H. (2021).
\newblock Birnbaum-saunders quantile regression and its diagnostics with
  application to economic data.
\newblock {\em Applied Stochastic Models in Business and Industry}, 37:53--73.

\bibitem[Saulo et~al., 2021]{ssls:21}
Saulo, H., Dasilva, A., Leiva, V., S\'anchez, L., and Fuente-Mella, H.~L.
  (2021).
\newblock Log-symmetric quantile regression models.
\newblock {\em Statistica Neerlandica}, DOI: 10.1111/stan.12243:1--26.

\bibitem[Schwarz, 1978]{s:78}
Schwarz, G. (1978).
\newblock {Estimation of the dimension of the model}.
\newblock {\em The Annals of Statistics}, 6:461--464.

\bibitem[Smithson and Shou, 2017]{Smithson17}
Smithson, M. and Shou, Y. (2017).
\newblock {CDF}-quantile distributions for modelling random variables on the
  unit interval.
\newblock {\em British Journal of Mathematical and Statistical Psychology},
  70(3):412--438.

\bibitem[Stasinopoulos and Rigby, 2007]{sta07}
Stasinopoulos, D. and Rigby, R. (2007).
\newblock Generalized additive models for location scale and shape (gamlss) in
  r.
\newblock {\em Journal of Statistical Software}, 23.

\bibitem[Tariq et~al., 2021]{Tariq:21}
Tariq, A., Undurraga, E., Laborde, C., Vogt-Geisse, K., and Luo, R. e.~a.
  (2021).
\newblock {Transmission dynamics and control of COVID-19 in Chile,
  March-October, 2020}.
\newblock {\em PLOS Neglected Tropical Diseases}, 15:e0009070.

\bibitem[Vitenu-Sackey and Barfi, 2021]{sackeybarfi:2021}
Vitenu-Sackey, P.~A. and Barfi, R. (2021).
\newblock The impact of {C}ovid-19 pandemic on the global economy: Emphasis on
  poverty alleviation and economic growth.
\newblock {\em The Economics and Finance Letters}, 81(0):32--43.

\bibitem[Waldmann, 2018]{waldmann:18}
Waldmann, E. (2018).
\newblock Quantile regression: a short story on how and why.
\newblock {\em Statistical Modelling}, 18.

\bibitem[Weisberg, 2014]{weisberg:14}
Weisberg, S. (2014).
\newblock {\em Applied Linear Regression}.
\newblock John Wiley \& Sons, Hoboken, New Jersey, fourth edition edition.

\end{thebibliography}

\end{document}